\documentstyle[multicol,aps,psfig]{revtex}
\begin{document}
%\draft
\preprint{LBNL-47155}
%\baselineskip=14pt
%\parindent=0.25in
%\abovedisplayskip=14pt
%\belowdisplayskip=14pt
%\parindent=0.5in
%LBNL-39408
%Sept. 24, 1996
%
%\renewcommand{\thefootnote}{\fnsymbol{footnote}}
%\setcounter{footnote}{0}

\title{Multiple Parton Scattering in Nuclei: Parton Energy Loss}
\author{Xin-Nian Wang}
\address{Nuclear Science Division, Mailstop 70-319,\\
Lawrence Berkeley National Laboratory, Berkeley, CA 94720 USA}
\author{Xiaofeng Guo}
\address{Department of Physics and Astronomy, University of Kentucky,\\
Lexington, Kentucky KY 40506,USA}

\date{February 17, 2001}

\maketitle

\vspace{-2.0in}
{\hfill LBNL-47155}
\vspace{2.in}

\begin{abstract}
\baselineskip=12pt
Multiple parton scattering and induced parton energy loss are
studied in deeply inelastic scattering (DIS) off nuclei.
The effect of multiple scattering of a highly off-shell 
quark and the induced parton energy loss is expressed 
in terms of the modification to the quark fragmentation functions. 
We derive such modified quark fragmentation functions and their QCD 
evolution equations in DIS using the generalized factorization of 
higher twist parton distributions. We consider double-hard and 
hard-soft parton scattering as well as their interferences in the same 
framework. The final result, which depends on both the diagonal and 
off-diagonal twist-four parton distributions in nuclei, demonstrates clearly 
the Landau-Pomeranchuk-Migdal interference features and predicts a unique
nuclear modification of the quark fragmentation functions.   

\end{abstract}

\pacs{ 24.85.+p, 12.38.Bx, 13.87.Ce, 13.60.-r}

\baselineskip=16pt

%\begin{multicols}{2}
%\end{multicols}

\section{Introduction}

Hard processes in high-energy strong interactions are always localized
in space-time because of the large momentum-energy transfer. The asymptotic
behavior of QCD allows one to compute these cross sections perturbatively.
Together with the factorization theorem and the experimental information of
parton distributions and fragmentation functions, hard processes in
hadronic collisions have been well understood \cite{pQCD}. One can then in
turn use them as probes of nuclear matter as well as hot quark-gluon
plasma which is expected to be formed in high-energy nuclear collisions.
In particular, the propagation of an energetic parton
and its induced energy loss has been proposed as a probe of the properties
of dense matter formed in high-energy nuclear collisions \cite{GP,GPTW,WG92}.
Based on a model of multiple scattering and induced radiation in QCD proposed
by Gyulassy and Wang (GW)\cite{GW1}, recent theoretical 
studies \cite{GW1,BDPS,BGZ,GLV,wied} show that a fast 
parton will lose a significant amount of energy via 
induced radiation when it propagates through a hot partonic matter. 
The most interesting feature of the result is the quadratical distance
dependence of the total energy loss because of the
non-Abelian nature of QCD radiation and the Landau-Pomercanchuk-Migdal (LPM)
\cite{LPM} interference. Such a quadratical distance
dependence is also a consequence of the GW static color-screened potential
model for multiple scattering where colors are screened but not confined.
In this paper we will study parton multiple scattering inside
a nucleus where colors are confined to the size of a nucleon. In this
case parton propagation will certainly be different from that in a partonic 
matter and one should expect that the parton energy loss to be related to
the nucleon size or confinement scale. Parton energy loss in $eA$ DIS
has been studied before within various models of intranuclear scattering
and the modification of hadronization \cite{NAG,niko,bialas,gp_eA,BZK}.
In this paper we will study the problem in the framework of multiple
parton scattering in perturbative QCD (pQCD).

Unlike the situation in QED, the energy loss of a parton cannot be 
directly measured because partons are not the final experimentally observed 
particles. The total energy of a jet as traditionally defined by a cluster of 
hadrons in the phase space will not change much due to medium induced 
radiation because a jet so defined contains particles both from the 
leading parton and from the radiated gluons. 
This is particularly the case if multiple 
scattering and induced radiation do not {\it dramatically } change the 
energy profile of the jet in phase-space. It is also virtually impossible
to determine the jet energy event by event because of the large background 
and its fluctuation in heavy-ion collisions. One then has to resort to 
particle distributions within a jet and study the effect of parton energy loss 
by measuring the modification of the particle distribution due to multiple 
scattering and induced radiation \cite{WG92}. One such distribution 
is the fragmentation function of the produced parton, 
$D_{a\rightarrow h}(z,\mu^2)$, where $z$ is the fractional 
energy of the parton $a$ carried by the produced particles $h$.
Unlike the situation that has been considered in most of the recent 
theoretical studies \cite{GW1,BDPS,BGZ} of parton energy loss of an 
on-shell parton during its propagation through medium, partons produced 
in hard processes are normally far off-shell characterized by the 
momentum scale $\mu^2$.  Final-state
radiation of these off-shell partons in free space, such as in $e^+e^-$ 
annihilation, leads to the $\mu$-dependence of the fragmentation functions as
given by Dokshitzer-Gribov-Lipatov-Altarelli-Parisi (DGLAP) \cite{dglap} QCD 
evolution equations,
\begin{eqnarray}
  \frac{\partial D_{q\rightarrow h}(z_h,\mu^2)}{\partial \ln \mu^2} & = &
  \frac{\alpha_s(\mu^2)}{2\pi} \int^1_{z_h} \frac{dz}{z} 
\left[ \gamma_{q\rightarrow qg}(z)
D_{q\rightarrow h}(z_h/z,\mu^2) + \gamma_{q\rightarrow gq}(z) 
D_{g\rightarrow h}(z_h/z,\mu^2)\right],  \label{eq:ap1} \\
\frac{\partial D_{g\rightarrow h}(z_h,\mu^2)}{\partial \ln \mu^2} & = &
\frac{\alpha_s(\mu^2)}{2\pi} \int^1_{z_h} \frac{dz}{z} \left[
    \sum_{q=1}^{2n_f} \gamma_{g\rightarrow q\bar{q}}(z)
  D_{q\rightarrow h}(z_h/z,\mu^2) + \gamma_{g\rightarrow gg}(z)
 D_{g\rightarrow h}(z_h/z,\mu^2)\right] \label{eq:ap2}
\end{eqnarray}
where $\gamma_{a\rightarrow bc}(y)$ 
are the splitting functions of the corresponding
radiative processes \cite{field}. 
When a parton is produced in a medium, it will suffer multiple scattering and
induced radiation that will then lead to medium modification of the DGLAP 
evolution of the parton fragmentation functions. In this paper we will 
derive such modified evolution equations for parton fragmentation 
functions in the simplest case of deeply inelastic $eA$ scattering (DIS). 
Multiple scattering and induced 
radiation suffered by the leading quark in this case will give rise to an
additional term in the DGLAP evolution equations. As a consequence, 
the modified fragmentation functions become softer. This can be 
directly translated into the energy loss of the leading quark.

The study here is very similar to that by Luo, 
Qiu and Sterman (LQS) \cite{LQS} 
on nuclear dependence of jet cross section in deeply inelastic 
scattering. The difference is that they considered large transverse momentum
jet production whereas we will concentrate on 
soft gluon emission which is responsible for
the DGLAP evolution equations of fragmentation functions. Depending on the 
fractional momentum carried by the second parton in the case of double parton 
scattering, one can categorize the processes into soft-hard, hard-hard and 
their corresponding interferences. In their study of large transverse momentum 
jet production, LQS only considered soft-hard and hard-hard processes, because
the interference terms become negligible as we will show. Since we have to 
consider soft gluon emission, the interference terms are very important 
and they will cancel contributions from hard-hard and hard-soft 
processes in the limit of collinear emission (or zero transverse momentum). 
In this paper, we will include all these processes and treat them in 
the same manner. Utilizing the generalized factorization of higher-twist 
parton distributions, we find that
each process in the double scattering probes different twist-four parton 
correlation of the nuclear medium. As a result of the sum of all contributions,
the additional term in the modified DGLAP evolution equation is proportional 
to the combined twist-four matrix element,
\begin{eqnarray}
 \frac{1 }{4\pi} \int & dy^- dy^-_1 dy^-_2 \theta(-y^-_2)\theta(y^--y^-_1)
 (1-e^{ix_Lp^+(y^-_1-y^-)})(1-e^{-ix_Lp^+y^-_2}) \nonumber \\
& e^{i(x_B+x_L)p^+y^-}
\langle A|\bar{\psi}_q(0)\gamma^+ F^{\alpha+}(y^-_2)
F_{\alpha}^+(y^-_1)\psi_q(y^-)
|A\rangle,
\end{eqnarray}
where
\begin{equation}
  x_L=\mu^2/2p^+q^-z(1-z),
\end{equation}
with $\mu$ the typical transverse momentum or factorization 
scale and $z$ the fractional momentum
of the emitted gluon. In this paper, we denote the four-momentum of the virtual
photon and the target nucleon in DIS as
\begin{eqnarray}
  q& = &[-Q^2/2q^-, q^-, \vec{0}_\perp], \label{eq:frame} \nonumber \\
  p& = &[p^+,0,\vec{0}_\perp],
\end{eqnarray}
respectively. The Bjorken variable is then $x_B=Q^2/2p^+q^-$. 

In the above matrix element, one can identify $1/x_Lp^+=2q^-z(1-z)/\mu^2$ 
as the formation time of the emitted gluons. For large formation time 
as compared to the nuclear size, the above matrix element vanishes,
demonstrating a typical LPM interference effect.
This is because the emitted gluon (with long formation time) and the leading 
quark are still a coherent system when they propagate through the nucleus.
Additional scattering will not induce more gluon radiation, 
thus limiting the energy loss of the leading quark. 

The reduction of phase space available for gluon radiation due to the 
LPM interference effect is critical for applying the LQS formalism to 
the problem in this paper. In the original LQS approach \cite{LQS}, the 
generalized factorization for processes with a large final transverse
momentum $\ell_T^2\sim Q^2$ allows one to consider the leading 
contribution in $1/Q^2$ which is enhanced by the nuclear 
size $R_A \sim A^{1/3}$. For large $Q^2$ and $A$, higher-twist contribution
from double parton rescattering that is proportional to $\alpha_s R_A/Q^2$
will then be the leading nuclear correction. One can neglect contributions 
from more than two parton rescattering. In deriving the modified fragmentation 
functions, we however have to take the leading logarithmic approximation 
in the limit $\ell_T^2\ll Q^2$, where $\ell_T$ is the transverse 
momentum of the radiated gluon. Since the LPM interference suppresses 
gluon radiation whose formation time ($\tau_f \sim Q^2/\ell_T^2p^+$)
is larger than the nuclear 
size $MR_A/p^+$ in our chosen frame, $\ell_T^2$ should then have a 
minimum value of $\ell_T^2\sim Q^2/MR_A\sim Q^2/A^{1/3}$. 
Here $M$ is the nucleon mass.
Therefore, the logarithmic approximation is still valid for
large nuclei ($MR_A\gg 1$). In the meantime, the leading higher-twist
contribution proportional to $\alpha_s R_A/\ell_T^2 \sim \alpha_s R_A^2/Q^2$
can still be small for large $Q^2$ so that one can neglect
processes with more than two parton rescattering. The parameter for
twist expansion of the fragmentation processes inside a nucleus is 
thus $\alpha_sA^{2/3}/Q^2$ as compared to $\alpha_sA^{1/3}/Q^2$ for
processes with large final transverse momentum as studied by LQS \cite{LQS}.
This is why the nuclear modification to the fragmentation function
as derived in this paper depends quadratically on the nuclear size $R_A$.

Since our approach allows us to relate the energy loss of a fast parton to 
the parton correlations in the medium, the final results will be sensitive 
to the properties of the medium through which a produced parton 
propagates. In particular, the medium-dependence of the above matrix 
elements in a deconfined quark-gluon plasma will be very different  
from that in an ordinary nuclear medium. Therefore, the parton energy 
loss and its dependence on the medium size in hadronic matter 
are expected to differ from that in a quark-gluon plasma.
This simply reflects the different parton correlations in 
different types of media.

The results of the present study were already reported in 
Ref.~\cite{xgxw00}. We will provide a detailed 
derivation and discussion in this paper.
The rest of this paper is organized as follows: In the next section, we give
a brief overview of the framework of our study including generalized
factorization of twist-four parton distributions in hard processes. 
In Section III we will discuss in detail the calculation of different 
contributions to soft gluon emission involving double parton scattering. 
To simplify the calculation, we will use the techniques of helicity 
amplitude with soft gluon approximation ($z<<1$). In Section IV, 
we will consider virtual corrections from unitary constraints. We will 
then define, in Section V, the effective parton fragmentation functions 
in deeply inelastic $eA$ collisions and derive the modified DGLAP evolution 
equations. We will also calculate the average energy loss suffered by 
the leading quark. Finally, in Section VI, we will discuss our results 
and their implications in other hard processes.

%%%%%%%%%%%%%%%%%%%%%%%%%%%%%%%%%%%%%%%%

\section{General Formalism}

Consider the following semi-inclusive process in the deeply inelastic 
lepton-nucleus scattering, 
\begin{equation}
e(L_1) + A(p) \longrightarrow e(L_2) + h (\ell_h) +X \ ,
\label{process}
\end{equation}
where $L_1$ and $L_2$ are the four-momenta of the incoming 
and the outgoing leptons respectively,  
$p$ is the momentum per nucleon for the nucleus with the atomic 
number $A$, and $\ell_h$ is the observed hadron momentum. The momentum
of the virtual photon ($\gamma^*$) is $q=L_2-L_1$.

The differential cross section of semi-inclusive processes 
in DIS with an observed final state hadron $\ell_h$ can be expressed as
\begin{equation}
E_{L_2}E_{\ell_h}\frac{d\sigma_{\rm DIS}^h}{d^3L_2d^3\ell_h}
=\frac{\alpha^2_{\rm EM}}{2\pi s}\frac{1}{Q^4} L_{\mu\nu}
E_{\ell_h}\frac{dW^{\mu\nu}}{d^3\ell_h} \; ,
\label{sigma-dis}
\end{equation}
where $s=(p+L_1)^2$ is the total invariant mass of the lepton-nucleon
system. The leptonic tensor $L_{\mu\nu}$ is given by  
\begin{equation}
L_{\mu\nu}=\frac{1}{2}\, {\rm Tr}(\gamma \cdot L_1 \gamma_{\mu}
\gamma \cdot L_2 \gamma_{\nu}) \; .
\end{equation}
The semi-inclusive hadronic tensor $E_{\ell_h}dW^{\mu\nu}/d^3\ell_h$ is
defined as,
\begin{equation}
E_{\ell_h}\frac{dW^{\mu\nu}}{d^3\ell_h}=
\frac{1}{2}\sum_X \langle A|J^\mu(0)|X,h\rangle 
\langle X,h| J^\nu(0)|A\rangle
2\pi \delta^4(q+p-p_X-\ell_h) \; ,
\end{equation}
where $\sum_X$ runs over all possible intermediate states and $J_\mu$ is the
hadronic electromagnetic (EM) current, 
$J_\mu=e_q \bar{\psi}_q \gamma_\mu\psi_q$.

\begin{figure}
\centerline{\psfig{file=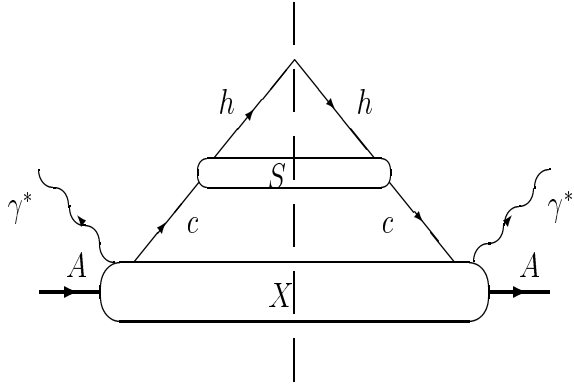,width=3in,height=2.0in}}
\caption{ Diagram representing the factorized form of the semi-inclusive 
hadronic tensor $W_{\mu\nu}$.}
\label{fig1}
\end{figure}

In the parton model with collinear factorization approximation, 
one can in general factorize the semi-inclusive cross section into a 
parton fragmentation function and the partonic cross section. By factorizing
the parton fields out of the EM current as
\begin{equation}
  J^\mu(0)\equiv \sum_{c,\alpha}\widetilde{J}^\mu_{c,\alpha}(0) 
  \hat{\phi}^{\alpha}_c(0) \equiv \sum_{c,\alpha} 
  \hat{\phi}^{\alpha\dagger}_c(0)\widetilde{J}^{\mu\dagger}_{c,\alpha}(0)\; ,
\end{equation}
one can express the semi-inclusive hadronic tensor as
\begin{equation}
\frac{dW^{\mu\nu}}{dz_h}
=\sum_{\alpha,\beta,c}z^2_h \hat{d}^{\alpha\beta}_{c\rightarrow h}(z_h,\ell_h)
\sum_X \langle A|\widetilde{J}^\mu_{c,\alpha}(0)|X\rangle 
\langle X| \widetilde{J}^{\nu\dagger}_{c,\beta}(0)|A\rangle
2\pi \delta(\ell_c^2) \; ,
\end{equation}
as illustrated by Fig.~\ref{fig1},
where $\alpha,\beta$ are indices (spinor for quarks and Lorentz for gluons)
of the partonic field of species $c$ with momentum $\ell_c$,  and 
$\hat{d}^{\alpha\beta}_{c\rightarrow h}(z_h,\ell_h)$, related to the 
fragmentation function, is defined as
\begin{equation}
  \hat{d}^{\alpha\beta}_{c\rightarrow h}(z_h,\ell_h)\equiv\sum_S 
\int\frac{d^4\ell_c}{(2\pi)^4} d^4y \langle 0| \hat{\phi}^\alpha_c(0)
|h,S\rangle\langle h,S|\hat{\phi}^{\beta\dagger}_c(y)|0\rangle
e^{-i\ell_c\cdot y} \delta(z_h - \frac{\ell_h^-}{\ell_c^-}). \label{eq:smd}
\end{equation}
After such factorization, one only needs to calculate the cross sections of
partonic hard processes. The fragmentation functions defined in terms of
parton matrix elements in the above equation are nonperturbative and can
only be obtained via experimental measurements. In addition, one can also
factorize out the parton distributions inside nuclei (or nucleons). Such
factorization has been proven to all orders and in the leading 
twist \cite{factorize}.

In the infinite-momentum frame as defined by Eq.~(\ref{eq:frame}), 
the dominant 
momentum component of the leading parton in a single jet event is the 
minus component. Therefore, we define the momentum fraction carried by 
a produced hadron as $z_h=\ell_h^-/\ell_c^-$. 

\subsection{Leading Twist Contributions}

\begin{figure}
\centerline{\psfig{file=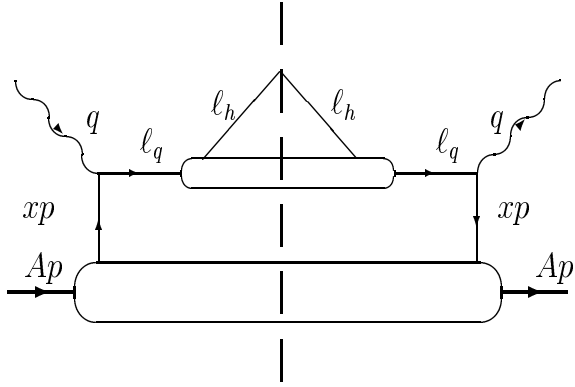,width=3in,height=2.0in}}
\caption{Lowest order process that contributes to $H^(0)_{\mu\nu}$.}
\label{fig2}
\end{figure}

The leading-twist contribution to DIS to the lowest order 
(${\cal O}(\alpha_s^0)$) comes from a single hard $\gamma^*+ q$ scattering
as illustrated in Fig.~\ref{fig2}. The semi-inclusive hadronic tensor
$dW_{\mu\nu}/dz_h$ can be factorized as \cite{factorize}
\begin{equation}
\frac{dW_{\mu\nu}^{S(0)}}{dz_h}
= \sum_q \int dx f_q^A(x) H^{(0)}_{\mu\nu}(x,p,q) D_{q\rightarrow h}(z_h)\ ,
\label{eq:w-s}
\end{equation}
where $f_q^A(x)$ is the quark distribution as defined by
\begin{eqnarray}
f_q^A(x)&\equiv&\int \frac{d^4k}{(2\pi)^4} d^4y e^{ik\cdot y}
\delta(x-\frac{k^+}{p^+}) 
\langle A| \bar{\psi}_q(0)\frac{\gamma^+}{2p^+}\psi_q(y)|A\rangle \nonumber \\
&=&\int \frac{dy^-}{2\pi}e^{ixp^+y^-} \frac{1}{2}
\langle A| \bar{\psi}_q(0)\gamma^+\psi_q(y^-)|A\rangle \; . \label{eq:fquark}
\end{eqnarray}
Here quark fragmentation function $D_{q\rightarrow h}(z_h)$ is defined as
\begin{eqnarray}
D_{q\rightarrow h}(z_h)&\equiv&\frac{z_h^3}{2}d_{q\rightarrow h}(z_h,\ell_h)
\equiv \frac{z_h^3}{4\ell_h^-} 
{\rm Tr}[\gamma^- \hat{d}_{q\rightarrow h}(z_h,\ell_h)], \label{eq:coll} \\
\hat{d}_{q\rightarrow h}^{\alpha\beta}(z_h,\ell_h)&\equiv&
\sum_S \int\frac{d^4\ell_q}{(2\pi)^4} d^4y e^{-i\ell_q\cdot y} 
\delta(z_h - \frac{\ell_h^-}{\ell_q^-})
\langle 0|\psi^\beta_q(0)|h,S\rangle\langle h,S|\bar{\psi}^\alpha_q(y)|0\rangle
\nonumber \\
&=& \sum_S \frac{\ell_h^-}{z_h^2} \int \frac{dy^+}{2\pi}
e^{-i\ell_h^-y^+/z_h}
\langle 0|\psi^\beta_q(0)|h,S\rangle\langle h,S
|\bar{\psi}^\alpha_q(y^+)|0\rangle \; .
\end{eqnarray}
The hard part of $\gamma^* +q$ partonic scattering is
\begin{eqnarray}
H_{\mu\nu}^{(0)} &=& e_q^2\frac{1}{2}\, 
{\rm Tr}(\gamma \cdot p \gamma_{\mu} \gamma \cdot(q+xp) \gamma_{\nu})
\, (2\pi) \delta[(q+xp)^2] \,
\nonumber \\ 
&=& 4\pi e_q^2\left[x_B e^L_{\mu\nu} - \frac{1}{2}e^T_{\mu\nu}\right]\delta (x-x_B) \, ,
\label{Hs-0}
\end{eqnarray}
where momentum conservation gives $\ell_q=xp+q$ and by definition
$\ell_h^-=z_h\ell_q^-=z_hq^-$. The transverse and longitudinal tensors 
are defined as
\begin{eqnarray}
  e^T_{\mu\nu}& = &g_{\mu\nu}-\frac{q_\mu q_\nu}{q^2}, \nonumber \\
  e^L_{\mu\nu}& = &\frac{1}{p\cdot q} 
[p_\mu-\frac{p\cdot q}{q^2} q_\mu]
[p_\nu-\frac{p\cdot q}{q^2} q_\nu] \, .
\label{eq-tensor}
\end{eqnarray}
Note that a sum and an average over color indices are implied
in the quark distribution and fragmentation function, respectively.
In Eq.~(\ref{eq:coll}), we used the collinear 
approximation in the unpolarized fragmentation functions,
\begin{eqnarray}
  \hat{d}_{q\rightarrow h}(z_h,\ell_h)&\approx& \frac{1}{2} 
  d_{q\rightarrow h}(z_h,\ell_h) \gamma\cdot \ell_h +\cdots, \nonumber \\
  &\approx& \frac{z_h}{2}  
  d_{q\rightarrow h}(z_h,\ell_h) \gamma\cdot \ell_q +\cdots,
\end{eqnarray}
while other higher-twist terms are neglected. We will use such a collinear
approximation throughout this paper even in the case of double scattering. We
will also neglect other higher-twist contributions to the fragmentation 
function that are independent of the nuclear size.

\begin{figure}
\centerline{\psfig{file=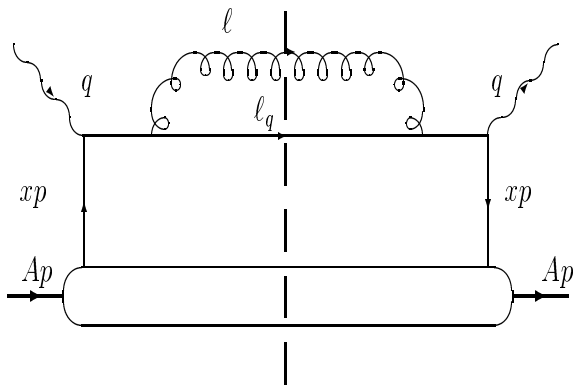,width=3in,height=2.0in}}
\caption{The hard partonic part of next-to-leading order process that
contributes to $H^{(1)}_{\mu\nu}$.}
\label{fig3}
\end{figure}

At the next-to-leading order ${\cal O}(\alpha_s)$, the dominant 
(in leading log approximation) real radiative contribution to the 
fragmentation process in an axial gauge ($A^-=0$) comes 
from the final state radiation. Since we are only interested
in the collinear behavior of the radiative corrections in order to study
the evolution equation of the fragmentation functions, we will keep only
the leading (divergent) contribution when the gluon's transverse momentum
vanishes(with respect to $q+xp$), {\it i.e.}, $\ell_T\rightarrow 0$. 
Using the same collinear approximation, one finds the leading 
contribution from the radiative correction
(for $\ell_T^2$ up to a factorization scale $\mu^2$) 
to the quark fragmentation process,
\begin{equation}
\frac{dW^{S(1)q}_{\mu\nu}}{dz_h}
= \sum_q \int dx f_q^A(x) \int_{z_h}^1 \frac{dz}{z}\,
D_{q\rightarrow h}(z_h/z)\, H^{(1)q}_{\mu\nu}(x,p,q,z)\, , \label{eq:h11-0}
\end{equation}
where $z_h=\ell_h^-/q^-$ and $z=\ell_q^-/q^-$ are the fractional 
momentum carried by hadron and the final quark, respectively.
The factorized form in Eq.~(\ref{eq:h11-0}) is quite general for all processes 
involving gluon radiation in the final state. It is then enough to calculate 
the partonic hard part $H^{(1)q}_{\mu\nu}(x,p,q,z)$ as illustrated
in Fig.~\ref{fig3},
\begin{equation}
H^{(1)q}_{\mu\nu}(x,p,q,z)=H^{(0)}_{\mu\nu}(x,p,q) \int_0^{\mu^2} 
\frac{d\ell_T^2}{\ell_T^2} \frac{\alpha_s}{2\pi}
C_F \frac{1+z^2}{1-z}\, , \label{eq:h11}
\end{equation}
 where the color factor is $C_F=(N_c^2-1)/2N_c=4/3$. Since we neglect the 
non-leading-log terms, the tensor structure of $H^{(1)q}_{\mu\nu}$ remains the
same as $H^{(0)}_{\mu\nu}$ in Eq. (\ref{Hs-0}).
Similarly, if the final hadron comes from the gluon fragmentation, 
the contribution is
\begin{equation}
\frac{dW^{S(1)g}_{\mu\nu}}{dz_h}
=\sum_q \int dx f_q^A(x) H^{(0)}_{\mu\nu}(x,p,q) \int_0^{\mu^2} 
\frac{d\ell_T^2}{\ell_T^2} \frac{\alpha_s}{2\pi} \int_{z_h}^1 \frac{dz}{z}\,
C_F \frac{1+(1-z)^2}{z} D_{g\rightarrow h}(z_h/z) \; . \label{eq:h12}
\end{equation}
Here $z=\ell_g^-/q^-$ and $D_{g\rightarrow h}(z_h)$ 
is the gluon fragmentation function defined as
\begin{eqnarray}
D_{g\rightarrow h}(z_h)&\equiv&\frac{z_h^2}{2}\varepsilon_{\mu\nu}(\ell_g)
d^{\mu\nu}_{g\rightarrow h}(z_h,\ell_h) \nonumber \\
&=&-\frac{z_h^2}{2\ell_h^-} \sum_S\int \frac{dy^+}{2\pi}
e^{-i\ell_h^-y^+/z_h} \langle 0 |F^{-\mu}(0)|S,h\rangle
\langle S,h|F^-_{\;\;\;\mu}(y^+)|0\rangle \, ,
\end{eqnarray}
where $d^{\mu\nu}_{g\rightarrow h}(z_h,\ell_h)$ is 
given by Eq.~(\ref{eq:smd}) for gluon fields and 
\begin{equation}
\varepsilon_{\mu\nu}(\ell_g)=
\sum_{\lambda=1,2}\varepsilon_\mu(\ell_g,\lambda)
\varepsilon_\nu(\ell_g,\lambda)\; ,
\end{equation}
with $\varepsilon_\mu(\ell_g,\lambda)$ being the polarization vector of a
gluon in an axial gauge.

There are both infrared and collinear divergences in Eq.~(\ref{eq:h11}) 
and Eq.~(\ref{eq:h12}). The infrared divergences come from the phase space
where the gluon's fractional momentum goes to zero. These divergences will
be canceled by the virtual corrections in the quark self-energy diagrams as
we will discuss later. The
collinear divergences when $\ell_T\rightarrow 0$ will be absorbed into the
renormalized fragmentation functions and are responsible for the 
evolution of the quark fragmentation function with respect to the
factorization scale $\mu$. Summing up all the contributions from 
Eqs.~(\ref{eq:w-s}), (\ref{eq:h11-0}) and (\ref{eq:h12}), one has
\begin{eqnarray}
\frac{dW^S_{\mu\nu}}{dz_h}
&=& \sum_q \int dx f_q^A(x) H^{(0)}_{\mu\nu}(x,p,q) 
D_{q\rightarrow h}(z_h,\mu^2)\, , \\
D_{q\rightarrow h}(z_h,\mu^2)&\equiv&D_{q\rightarrow h}(z_h)+
\int_0^{\mu^2} \frac{d\ell_T^2}{\ell_T^2} 
\frac{\alpha_s}{2\pi} \int_{z_h}^1 \frac{dz}{z}
\left[ C_F\frac{1+z^2}{1-z} D_{q\rightarrow h}(z_h/z)\right. \nonumber \\
&+& \left. C_F\frac{1+(1-z)^2}{z} D_{g\rightarrow h}(z_h/z)\right] 
+{\rm virtual\,\, corrections}\, ,
\label{eq:s-sum}
\end{eqnarray}
where the renormalized quark fragmentation function 
$D_{q\rightarrow h}(z_h,\mu^2)$ satisfies the
DGLAP evolution equation in Eq.~(\ref{eq:ap1})
after inclusion of the virtual corrections. The virtual corrections have 
to be included to ensure unitarity and infrared safety of the final result.
We will discuss the virtual contributions later in this paper.

\subsection{Higher Twist Contributions}
\label{secii}

In a nuclear medium, the outgoing quark in DIS may experience additional 
scattering with other partons from the nucleus. The additional scattering
may induce additional gluon radiation and cause the leading quark to 
lose energy. Such induced gluon radiation will effectively give 
rise to additional terms in the evolution equation leading to modification 
of the fragmentation functions in a medium. These additional terms 
from multiple scattering are always non-leading-power contributions. 
There are many next-leading-twist contributions to the semi-inclusive
hadronic tensor in DIS. One type of contribution involves non-leading-twist
matrix elements of parton distributions in the nucleus which in 
general are related to two-parton correlations. Since two-parton 
correlations in a nucleus can involve partons from different nucleons in the 
nucleus, they are proportional to the size of the nucleus and thus are 
enhanced by a nuclear factor $A^{1/3}$ as compared to two-parton correlations
in a nucleon. In this paper we will consider only those non-leading-twist
contributions that are enhanced by the nuclear factor. We will neglect 
those contributions that are not enhanced by the nuclear medium. 
They are in general not related to multiple parton scattering in the nucleus.

\begin{figure}
\centerline{\psfig{file=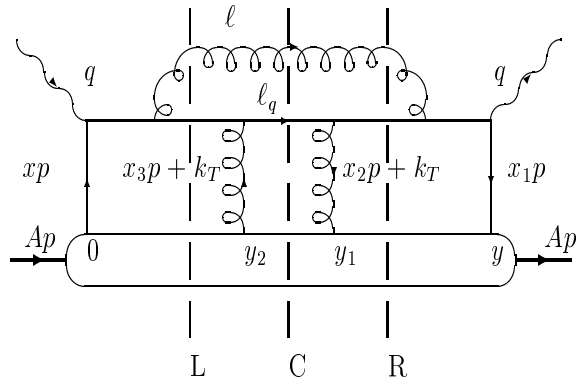,width=3in,height=2.0in}}
\caption{A diagram for quark-gluon rescattering processes with three possible
  cuts, central(C), left(L) and right(R).}
\label{fig4}
\end{figure}

In the LQS generalized factorization scheme for processes in nuclei with
large final transverse momentum ($\ell_T^2\sim Q^2$), contributions
from double parton scattering are leading in $\alpha_s A^{1/3}/Q^2$.
One can similarly apply the generalized factorization to the fragmentation
processes in nuclei thanks to the LPM interference effect in gluon radiation
induced by multiple parton scattering. As we have argued in the Introduction
and will demonstrate later in this paper, the formation time of gluon
radiation due to the LPM interference requires the radiated gluon to have
a minimum transverse momentum $\ell_T^2\sim Q^2/MR_A\sim Q^2/A^{1/3}$. 
The nuclear corrections to the fragmentation function due to double 
parton scattering will then be in the order of 
$\alpha_s A^{1/3}/\ell_T^2 \sim \alpha_s A^{2/3}/Q^2$. For large values of
$A$ and $Q^2$, these corrections are leading and yet the requirement
$\ell_T^2\ll Q^2$ for the logarithmic approximation in deriving the 
modified DGLAP evolution is still valid. Therefore, for large $Q^2$ and $A$, 
we can neglect processes with more than two parton rescattering and
consider only double scattering.

We will work in the LQS framework \cite{QS}
in which the twist-four parton distributions are factorizable. We will simply
apply such factorization of twist-four parton distributions to the study of 
semi-inclusive processes in DIS. In general, the twist-four contributions 
can be expressed as the convolution of partonic hard parts and 
four-parton matrix elements \cite{LQS}. At the lowest order 
(processes without gluon radiation) in this framework, rescattering 
with collinear gluons gives the eikonal contribution to the
gauge-invariant leading-twist and lowest-order result in Eq.~(\ref{eq:w-s}), 
assuming collinear factorization of the quark fragmentation function. 
As we will show later, rescattering with another quark in the leading order 
will contribute to the semi-inclusive cross section but not to the 
renormalization equation because there is no collinear divergency. 
Therefore, we first concentrate on processes involving rescattering 
with gluons. For next-to-leading order processes (with gluon radiation) 
involving a secondary scattering with another gluon from the nucleus, 
shown in Fig.~\ref{fig4} as an example, the double scattering 
contributions to $dW_{\mu\nu}/dz_h$ can be expressed as \cite{LQS} 
\begin{eqnarray}
\frac{dW_{\mu\nu}^D}{dz_h} &=& 
\sum_q \,\int_{z_h}^1\frac{dz}{z}D_{q\rightarrow h}(z_h/z)
\int \frac{dy^-}{2\pi}\, dy_1^-\, dy_2^-\,\frac{d^2y_T}{(2\pi)^2}
d^2k_T \overline{H}^D_{\mu\nu}(y^-,y_1^-,y^-_2,k_T,p,q,z); \nonumber \\
&\ \times & e^{- i\vec{k}_T \cdot\vec{y}_T}\frac{1}{2} \,\langle A |
 \bar{\psi}_q(0)\, \gamma^+ \,A^+(y_{2}^{-},0_{T})\,
          A^+(y_{1}^{-},y_{T})\, \psi_q(y^{-}) |A\rangle \ .
\label{WD1}
\end{eqnarray}
Here $\overline{H}^D_{\mu\nu}(y^-,y_1^-,y^-_2,k_T,p,q,z)$ is the
Fourier transform of the partonic hard part 
$\widetilde{H}_{\mu\nu}(x,x_1,x_2,k_T,p,q,z)$ in momentum space,
\begin{eqnarray}
\overline{H}^D_{\mu\nu}(y^-,y_1^-,y^-_2,k_T,p,q,z) 
&=& \int dx\, \frac{dx_1}{2\pi}\, \frac{dx_2}{2\pi}\, 
       e^{ix_1p^+y^- + ix_2p^+y_1^- + i(x-x_1-x_2)p^+y_2^-}\nonumber \\
&\ & \times \widetilde{H}^D_{\mu\nu}(x,x_1,x_2,k_T,p,q,z)\ ,
\label{eq-mH}
\end{eqnarray}
where $k_T$ is the relative transverse momentum carried by the second parton
in the double scattering. Values of the momentum fractions $x, x_1,$ and $x_2$ 
are fixed by $\delta$-functions and poles in the partonic hard part. 
They normally depend on $k_T$.

In order to pick up the next-leading-twist contribution, we 
expand the partonic hard part around $k_T=0$,
\begin{eqnarray}
\overline{H}^D_{\mu\nu}(y^-,y_1^-,y^-_2,k_T,p,q,z)
&=& \overline{H}^D_{\mu\nu}(y^-,y_1^-,y^-_2,k_T=0,p,q,z) \nonumber \\
&+& \left. \frac{\partial \overline{H}^D_{\mu\nu}}{\partial k_{T}^{\alpha}}
    \right|_{k_{T}=0}\ k_{T}^{\alpha}\, 
+\, \left. \frac{1}{2}\, \frac{\partial^{2}\overline{H}^D_{\mu\nu}}
                {\partial k_{T}^{\alpha} \partial  k_{T}^{\beta}} 
    \right|_{k_{T}=0}\ k_{T}^{\alpha}\, k_{T}^{\beta} + \ldots\ .
\label{eq-expand1}
\end{eqnarray}
This is known as collinear expansion \cite{qiu},
On the right-hand-side of Eq.~(\ref{eq-expand1}), the first term gives
the eikonal contribution to the leading-twist results. It does not 
correspond to the physical double scattering, but simply makes the 
matrix element in a single scattering gauge invariant. The second term for 
unpolarized initial and final states vanishes after being integrated 
over $k_T$.  The third term will give a finite contribution to the double 
scattering process. Substituting Eq.~(\ref{eq-expand1}) into
Eq.~(\ref{WD1}), and integrating over $d^2k_T$ and $d^2y_T$, we obtain
\begin{eqnarray}
\frac{dW_{\mu\nu}^D}{dz_h}
&=& \sum_q \,\int_{z_h}^1\frac{dz}{z}D_{q\rightarrow h}(z_h/z)
   \int \frac{dy^{-}}{2\pi}\, dy_1^-dy_2^-
\frac{1}{2}\, 
     \langle A | \bar{\psi}_q(0)\,
           \gamma^+\, F_{\sigma}^{\ +}(y_{2}^{-})\, 
 F^{+\sigma}(y_1^{-})\,\psi_q(y^{-})
     | A\rangle \nonumber \\
&\times &
    \left(-\frac{1}{2}g^{\alpha\beta}\right)
\left[\, \frac{1}{2}\, \frac{\partial^{2}}
                   {\partial k_{T}^{\alpha} \partial k_{T}^{\beta}}\,
    \overline{H}^D_{\mu\nu}(y^-,y_1^-,y^-_2,k_T,p,q,z)\, \right]_{k_T=0}\, ,
\label{eq-expand2}
\end{eqnarray}
where $k_T^{\alpha}A^+k_T^{\beta}A^+$ are converted into field strength 
$F^{\alpha+}F^{\beta+}$ by partial integrations.

\section{Quark-gluon double scattering}

In this section we calculate the hard part of parton rescattering with gluons.
We consider processes with a quark and a gluon in the initial state. In these
processes, there is first a hard photon-quark scattering. The produced parton
from the first hard scattering then has a second scattering with another
initial gluon from the nucleus. We refer to such processes as quark-gluon
double scattering processes.
We defer the calculation of virtual corrections to the
latter part of this paper. There are a total of 23 cut diagrams 
corresponding to 3 different double scattering processes,
their interferences, and the interferences between single scattering
and 7 different triple scattering processes (see Appendix). 
We use the process shown in Fig.~\ref{fig4} with three different cuts 
as an example to outline the steps that are essential to evaluate the hard 
part of double scattering. 

Since we are interested in the leading log behavior of the fragmentation
functions, we will only retain the leading contributions in the limit
of $\ell_T\rightarrow 0$ ($\ell_T$ being gluons transverse momentum) and 
neglect contributions that are finite when $\ell_T=0$. This corresponds
to the leading log approximation for $\ell_T^2\ll Q^2$.

\subsection{Hard vs Soft Rescattering} 

Following the collinear approximation involving leading twist-4 parton 
distributions, as outlined in Ref.~\cite{LQS}, the hard partonic part
in Fig.~\ref{fig4} with a central cut is

\begin{eqnarray}
\overline{H}^D_{C\,\mu\nu}(y^-,y_1^-,y_2^-,k_T,p,q,z)&=&
\int dx\frac{dx_1}{2\pi}\frac{dx_2}{2\pi}
e^{ix_1p^+y^- + ix_2p^+y_1^- + i(x-x_1-x_2)p^+y_2^-}
\int \frac{d^4\ell}{(2\pi)^4} \nonumber \\
&\times&\frac{1}{2}{\rm Tr}\left[p\cdot\gamma\gamma_\mu p^\sigma p^\rho 
\widehat{H}_{\sigma\rho}\gamma_\nu \right] 
2\pi\delta_+(\ell^2)\,
\delta(1-z-\frac{\ell^-}{q^-}) \; . \label{eq:hat1}
\end{eqnarray}
Here,
\begin{eqnarray}
\widehat{H}_{\sigma\rho} &=&
\frac{C_F}{2N_c}g^4\frac{\gamma\cdot(q+x_1 p)}{(q+x_1p)^2-i\epsilon}
\,\gamma_\alpha\,\frac{\gamma\cdot(q + x_1 p -\ell)}{(q+x_1p-\ell)^2-i\epsilon}
\,\gamma_\sigma \gamma\cdot\ell_q\,\gamma_\rho
\nonumber \\
&\times &\varepsilon^{\alpha\beta}(\ell)\frac{\gamma\cdot(q+xp -\ell)}{(q+xp-\ell)^2+i\epsilon} \,\gamma_\beta\,
\frac{\gamma\cdot(q + xp)}{(q+xp)^2+i\epsilon} 
\,\,2\pi \delta_+(\ell_q^2)\, , \label{eq:hat2} 
\end{eqnarray}
and
\begin{equation}
\varepsilon^{\alpha\beta}(\ell)=-g^{\alpha\beta} + 
\frac{n^\alpha\ell^\beta+n^\beta\ell^\alpha}{n\cdot\ell}
- n^2 \frac{\ell^\alpha \ell^\beta}{(n\cdot\ell)^2}
\end{equation}
is the polarization tensor of a gluon propagator in an axial gauge
($n\cdot A=0$) with $n=[1,0^-,\vec{0}_\perp]$ and
$\ell$, $\ell_q=q+(x_1+x_2)p+k_T-\ell$
are the 4-momenta carried by the gluon and the final quark, respectively.
$z=\ell_q^-/q^-$ is the fraction of longitudinal momentum 
(the large minus component) carried by the final quark.

To simplify the calculation of the trace part of 
$\widehat{H}_{\sigma\rho}$ and extract the leading contribution
in the limit $\ell_T\rightarrow 0$ and $k_T\rightarrow 0$, we again
use the collinear approximation,
\begin{equation}
p^\sigma\widehat{H}_{\sigma\rho}p^\rho
\approx \gamma\cdot\ell_q \,\frac{1}{4\ell_q^-}
{\rm Tr} \left[\gamma^- p^\sigma\widehat{H}_{\sigma\rho}p^\rho\right] \; .
\label{eq:coll2}
\end{equation}

The two longitudinal components in the $\ell$-integration are fixed
by the two $\delta$-functions in Eq.~(\ref{eq:hat1}) and the remainder becomes
$d\ell_T^2/(4\pi)^2$. To carry out the integration over $x,x_1$ and $x_2$,
we rewrite the $\delta$-function in Eq.~(\ref{eq:hat2}) as
\begin{equation}
\delta_+(\ell_q^2) = \frac{1}{2p^+q^-z}\delta(x_1+x_2-x_L-x_D-x_B) \; ,
\end{equation}
where $x_B=Q^2/2p^+q^-$ is the Bjorken variable and
\begin{eqnarray}
  x_L&=&\frac{\ell_T^2}{2p^+q^-z(1-z)} \,\, ,\,\,
  x_D=\frac{k_T^2-2\vec{k}_T\cdot \vec{\ell}_T}{2p^+q^-z} \, .
\label{eq:xld}
\end{eqnarray}
With the help of the on-shell condition $\ell_q^2=0$ and $\ell^2=0$ 
in the $\delta$-functions, one can also
simplify the variables of the propagators in Eq.~(\ref{eq:hat2}):
\begin{eqnarray}
(q+xp)^2&=&2p^+q^-(x-x_B) \,\, , 
(q+xp-\ell)^2=2p^+q^-z(x-x_L-x_B) \, ,\nonumber \\
(q+x_1p)^2&=&2p^+q^-(x_1-x_B) \,\, , 
(q+x_1p-\ell)^2=2p^+q^-z(x_1-x_L-x_B)\, .
\end{eqnarray}
The integration over $x$, $x_1$ and $x_2$ can now be cast in the form
\begin{eqnarray}
I_C(y^-,y_1^-,y_2^-,\ell_T,k_T,p,q,z)&=&\int dx \frac{dx_1}{2\pi}\frac{dx_2}{2\pi}
\frac{e^{ix_1p^+y^- + ix_2p^+y_1^- + i(x-x_1-x_2)p^+y_2^-}}
{(x_1-x_B -i\epsilon)(x_1-x_L-x_B-i\epsilon)} \nonumber \\
\;\;\;\;&\cdot&\frac{\delta(x_1+x_2-x_L-x_D-x_B)}
{(x-x_L-x_B+i\epsilon)(x-x_B+i\epsilon)} \, .
\end{eqnarray}
Two of the above integrations can be carried out by contour integration.
There are four possible poles in the denominators and two of them are used
to obtain the residues of the contour integration. There are four 
possible choices for the pair of poles representing subprocesses 
with different kinematics. When we choose the poles
\begin{equation}
x = x_B+x_L\; , \; x_1 = x_B+x_L\, ,
\end{equation}
which is the momentum fraction carried by the initial quark, 
the momentum fraction carried by the initial gluon in the 
rescattering is $x_2 = x_D$, which vanishes 
when $k_T\rightarrow 0$ according to the definition in
Eq.~(\ref{eq:xld}). In this case, the final gluon is produced via the 
final state radiation of the hard photon-quark scattering. 
After the final state radiation, 
the quark becomes on-shell and encounters another scattering with a very soft 
gluon. This subprocess corresponds to rescattering with a soft gluon after a 
hard collision and is normally referred to as a hard-soft process\cite{LQS}.
On the other hand, if we choose the poles, 
\begin{equation}
x = x_B\; , \; x_1 = x_B ,
\end{equation} 
the momentum fraction carried by the initial gluon in the 
rescattering is $x_2 = x_L+x_D$,
which is finite ($x_L$) even when $k_T\rightarrow 0$. We call this
type of rescattering hard and refer the corresponding double scattering
as double-hard processes. According to such a choice of a pair of poles,
the quark becomes on-shell immediately after the hard photon-quark scattering.
The gluon radiation is actually induced by the hard secondary quark-gluon
scattering. It is thus produced by the initial state radiation 
of the secondary scattering. The other two combinations of the poles,
\begin{eqnarray}
x & =& x_B+x_L\; , \; x_1=x_B\; ; \nonumber \\
x_1&=& x_B+x_L\; , \; x=x_B,
\end{eqnarray}
represent the interferences between hard-soft and double-hard processes.
Using the relations imposed by the $\delta$-function, $x_2=x_B+x_L+x_D-x_1$,
and the momentum conservation, $x_3=x_1+x_2-x$, one can verify that
there is a momentum transfer $x_L$ between the initial quark and gluon
fields. If we consider the quark and gluon come from different nucleons
inside the nucleus, this means that there is a momentum transfer between
different nucleons in these interferences processes. Such an observation
will become important when we study later the parton matrix elements
involved.
  
One can easily find out the residues of the contour integrations with
the above four possible choices of poles. After making a change of variable
$x_1+x_2-x_L-x_D\rightarrow x$, one finds
\begin{eqnarray}
I_C(y^-,y_1^-,y_2^-,\ell_T,k_T,p,q,z)&=&\int dx  \frac{\delta(x-x_B)}{x_L^2} 
\overline{I}_C(y^-,y_1^-,y_2^-,\ell_T,k_T,x,p,q,z)\, ; \nonumber \\
\overline{I}_C(y^-,y_1^-,y_2^-,\ell_T,k_T,x,p,q,z)
&=&e^{i(x+x_L)p^+y^- + ix_Dp^+(y_1^- - y_2^-)}
\theta(-y_2^-)\theta(y^- - y_1^-) \nonumber \\
&\times &(1-e^{-ix_Lp^+y_2^-})(1-e^{-ix_Lp^+(y^- - y_1^-)}) \; .
\label{eq:res1}
\end{eqnarray}
In the above contributions, the interferences between hard-soft and 
double-hard processes have the negative sign relative to the hard-soft 
and double-hard contributions. Previous studies \cite{LQS,Guo} have
considered only hard-soft and double-hard processes,
but neglected the interferences. For large $\ell_T$ or $x_L$, 
this is justified because the interference terms
will vanish due to the oscillation of the phase factors in the integration
of $y_1^-$ and $y_2^-$ over the size of the whole nucleus. Note that in 
the double-hard term the phase factor $\exp[ix_Lp^+(y_1^- -y_2^-)]$ does 
not vanish after integration over $y_1^-$ and $y_2^-$, because color 
confinement requires that $|y_1^- - y_2^-|$ remain the size of a nucleon. 
We will discuss other consequences of color confinement later
in this paper. At the $\ell_T \rightarrow 0$ limit, 
one can no longer neglect the 
interference terms because they exactly cancel the contributions 
from the hard-soft and double-hard processes. 
This was pointed out in Ref.~\cite{Guo}.
However, our final results, for the first time 
explicitly demonstrate such cancellation.
In the latter part of this section, we will discuss the physical implications
of such cancellation and its connection with the
LPM interference effect. It will
help us to reorganize our final results.

Because of the collinear (or leading log) approximation, 
the tensor structure of the double scattering contributions 
is generally the same as in the leading order single scattering,
\begin{equation}
\overline{H}^D_{\mu\nu}(y^-,y_1^-,y_2^-,k_T,p,q,z) =
\int dx H^{(0)}_{\mu\nu}(x,p,q)\
\overline{H}^D(y^-,y_1^-,y_2^-,k_T,x,p,q,z)\, . \label{eq:hc0}
\end{equation}
After completing the trace in Eq.~(\ref{eq:coll2}) and using the results
of the contour integration in Eq.~(\ref{eq:res1}), we have the
contributions from Fig.~\ref{fig4} with a central cut,
\begin{eqnarray}
\overline{H}^D_C(y^-,y_1^-,y_2^-,k_T,x,p,q,z)&=&
\int \frac{d\ell_T^2}{\ell_T^2}\, \frac{\alpha_s}{2\pi}\,
 C_F\frac{1+z^2}{1-z} \nonumber \\
&\times&\frac{2\pi\alpha_s}{N_c} 
\overline{I}_C(y^-,y_1^-,y_2^-,\ell_T,k_T,x,p,q,z)
 \, , \label{eq:hc}
\end{eqnarray}
where $\overline{I}_C$ is given in Eq.~(\ref{eq:res1}) and $H^{(0)}_{\mu\nu}$
in Eq.~(\ref{Hs-0}).

In addition to the central-cut diagram, we also need to consider 
asymmetrical-cut diagrams in Fig.~\ref{fig4} that represent
interferences between single and triple scattering. 
The trace part is exactly the same as in the central-cut diagram. 
The differences are that one of the quark propagators is replaced by the 
cut propagator in the central-cut diagram. In addition, each of the
asymmetrical-cut diagrams has only two possible pairs of poles, 
whereas there are four choices in the central-cut diagram.
After similar contour integration, we have for the left-cut diagram
\begin{eqnarray}
I_L(y^-,y_1^-,y_2^-,\ell_T,k_T,p,q,z)
&=&\int dx \frac{dx_1}{2\pi}\frac{dx_2}{2\pi}
\frac{e^{ix_1p^+y^- + ix_2p^+y_1^- + i(x-x_1-x_2)p^+y_2^-}}
{(x_1-x_B-i\epsilon)(x_1-x_L-x_B-i\epsilon)} \nonumber \\
\;\;\;\;&\cdot&\frac{\delta(x-x_B-x_L)}
{(x_1+x_2-x_B-x_L-x_D-i\epsilon)(x-x_B +i\epsilon)} \nonumber \\
&=&\int dx  \frac{\delta(x-x_B)}{x_L^2} 
\overline{I}_L(y^-,y_1^-,y_2^-,\ell_T,k_T,x,p,q,z)\, ; \nonumber \\
\overline{I}_L(y^-,y_1^-,y_2^-,\ell_T,k_T,x,p,q,z)
&=&-e^{i(x+x_L)p^+y^- + ix_Dp^+(y_1^- - y_2^-)}
\theta(y_1^- - y_2^-)\theta(y^- - y_1^-) \nonumber \\
&\times &(1-e^{-ix_Lp^+(y^- - y_1^-)}) \, .\label{eq:res2}
\end{eqnarray}
In the second step of the above equation we have made a change of variable
$x-x_L\rightarrow x$. On the left-hand side of the left-cut diagram, 
the kinematics of the final state gluon 
radiation in the single hard photon-quark scattering
requires the quark line to be off-shell, {\it i.e.}, $x=x_B$. Similarly, 
on the right-hand side, only one of the quark propagators 
connecting the gluon radiation vertex 
can be on shell. This leads to two possible pair of poles
in the propagators,
\begin{eqnarray}
x_1&=&x_B+x_L\, ,\, x_2=x_D\, ; \nonumber \\
x_1&=&x_B\, ,\, x_2=x_L+x_D\, ,
\end{eqnarray}
giving rise to the two contributions in Eq.~(\ref{eq:res2}). They correspond 
to soft and hard first quark-gluon rescattering, respectively,
in the triple scattering process. In both cases,
$x_3=x_D$. So the second quark-gluon rescattering in the triple scattering
process on the right-hand side of the left-cut diagram is always soft.

Similarly, one can also obtain for the right-cut diagram of Fig.~\ref{fig4},
\begin{eqnarray}
\overline{I}_R(y^-,y_1^-,y_2^-,\ell_T,k_T,x,p,q,z)
&=&-e^{i(x+x_L)p^+y^- + ix_Dp^+(y_1^- - y_2^-)}
\theta(-y_2^-)\theta(y_2^- - y_1^-) \nonumber \\
&\times &(1-e^{-ix_Lp^+y_2^-}) \, .\label{eq:res3}
\end{eqnarray}
Other coefficients of the asymmetrical-cut diagrams are exactly
the same as in the central-cut diagram given in Eq.~(\ref{eq:hc}).

\begin{figure}
\centerline{\psfig{file=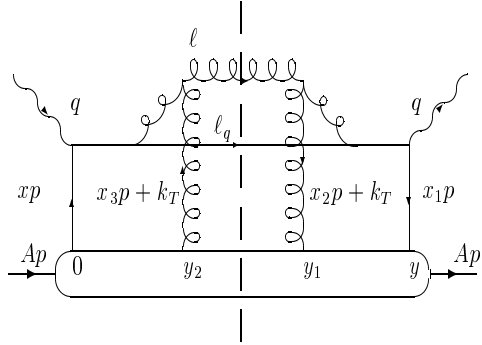,width=2.5in,height=1.8in}}
\caption{A diagram for the gluon-gluon rescattering process.}
\label{fig5}
\end{figure}

Since QCD is a non-Abelian gauge theory,  rescattering can also happen
between radiated and target gluons. Shown in Fig.~\ref{fig5}  
is an example that involves gluon-gluon rescattering. The contribution
to the hard part of double scattering from this central cut diagram is
\begin{eqnarray}
\overline{H}^D_{\rm Fig.\ref{fig5}}(y^-,y_1^-,y_2^-,k_T,x,p,q,z)&=&
\int \frac{d\ell_T^2}{(\vec{\ell_T}-\vec{k_T})^2}\, \frac{\alpha_s}{2\pi}\,
 C_A\frac{1+z^2}{1-z} \nonumber \\
&\times&\frac{2\pi\alpha_s}{N_c} 
\overline{I}_{\rm Fig.\ref{fig5}}(y^-,y_1^-,y_2^-,\ell_T,k_T,x,p,q,z)
 \, , \nonumber \\
\overline{I}_{\rm Fig.\ref{fig5}}(y^-,y_1^-,y_2^-,\ell_T,k_T,x,p,q,z)&=&
e^{i(x+x_L)p^+y^-+ix_Dp^+(y_1^--y_2^-)} 
\theta(-y_2^-)\theta(y^- - y_1^-) \nonumber \\
&\times&[e^{ix_Dp^+y_2^-/(1-z)}-e^{-ix_Lp^+y_2^-}] \nonumber \\
&\times&[e^{ix_Dp^+(y^- - y_1^-)/(1-z)}-e^{-ix_Lp^+(y^- - y_1^-)}] \, ,
\label{eq:hc2}
\end{eqnarray}
which has a structure very similar to the contribution in Eq.~(\ref{eq:res1}) 
and (\ref{eq:hc}) from the central-cut diagram in Fig.~\ref{fig4}. The four
different terms correspond similarly to the hard-soft, double-hard scattering
and their interferences. In the hard-soft process, 
the emitted gluon encounters 
another scattering with a very soft target gluon after 
it is produced in the final state radiation 
of the hard photon-quark scattering. Note that the initial
quark has momentum fraction $x=x_B+x_L+x_D/(1-z)$ while the soft gluon carries
momentum fraction $x_2=x_D-x_D/(1-z)$. In the double-hard process, 
the gluon is induced by the 
secondary hard scattering, similar to the quark-gluon
rescattering, but in this case via a three-gluon vertex. The $k_T$ dependence 
of the cross section is typical of the gluon-gluon interaction.

\begin{figure}
\centerline{\psfig{file=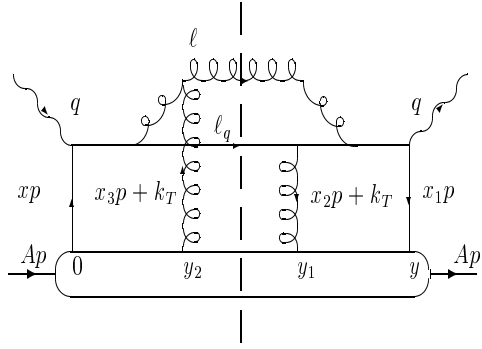,width=2.5in,height=1.8in}}
\caption{A diagram for the interference between quark-gluon and gluon-gluon
rescattering.}
\label{fig6}
\end{figure}

As another example, we illustrate the interference between quark-gluon and
gluon-gluon rescattering processes in Fig.~\ref{fig6}. The contribution
from the central-cut diagram is,
\begin{eqnarray}
\overline{H}^D_{\rm Fig.\ref{fig6}}(y^-,y_1^-,y_2^-,k_T,x,p,q,z)&=&
\int d\ell_T^2 \frac{\vec{\ell_T}\cdot(\vec{\ell_T}-\vec{k_T})}
{\ell_T^2(\vec{\ell_T}-\vec{k_T})^2}\, \frac{\alpha_s}{2\pi}\,
 \frac{C_A}{2}\frac{1+z^2}{1-z} \nonumber \\
&\times&\frac{2\pi\alpha_s}{N_c} 
\overline{I}_{\rm Fig.\ref{fig6}}(y^-,y_1^-,y_2^-,\ell_T,k_T,x,p,q,z)
 \, , \nonumber \\
\overline{I}_{\rm Fig.\ref{fig6}}(y^-,y_1^-,y_2^-,\ell_T,k_T,x,p,q,z)&=&
-e^{i(x+x_L)p^+y^- + ix_Dp^+(y_1^--y_2^-)} 
\theta(-y_2^-)\theta(y^- - y_1^-) \nonumber \\
&\times&[e^{ix_Dp^+y_2^-/(1-z)}-e^{-ix_Lp^+y_2^-}][1-e^{-ix_Lp^+(y^- - y_1^-)}]
\, .
\label{eq:hc3}
\end{eqnarray}
Notice that all interference terms in 
Eqs.~(\ref{eq:res2}-\ref{eq:res3}) and (\ref{eq:hc3}) 
have the opposite sign as compared to symmetrical diagrams.
The calculation of the other diagrams with all possible cuts is tedious 
but similarly straightforward. The splitting functions in all
quark-gluon double scattering and their interferences have the same form 
in $z$ as the $q\rightarrow qg$ splitting function in the single scattering 
case. The residues of the contour integrations in different processes are
different and the resultant different phase factors are essential to give 
rise to the interesting physical consequences as we discuss below.

\subsection{LPM Interference}
\label{sec:list}
\begin{figure}
\vspace{0.1in}
\centerline{\psfig{file=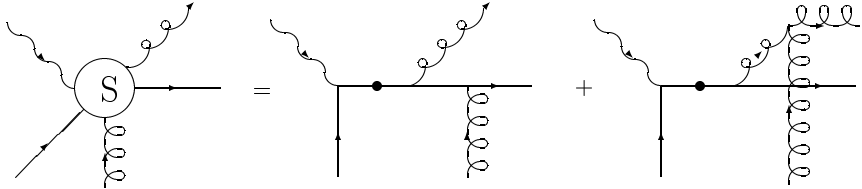,width=4.5in,height=1.0in}}
\caption{The diagrammatical representation of hard-soft processes. The
off-shell quark line is marked by a filled circle.}
\label{fig7}
\end{figure}

\begin{figure}
\centerline{\psfig{file=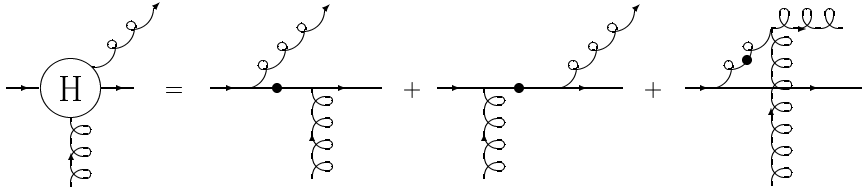,width=4.5in,height=1.0in}}
\caption{The diagrammatical representation of hard rescattering processes.
The off-shell parton lines are marked by a filled circle.}
\label{fig8}
\vspace{0.1in}
\end{figure}

Before we list the results of all processes, it is helpful to 
discuss first the underlying physics in gluon radiation induced by 
quark-gluon double scattering. As we have seen in the calculation of
the contributions from diagrams in Figs.~\ref{fig4}-\ref{fig6}, 
the so-called hard-soft processes correspond to the case where the 
gluon radiation is induced by the hard scattering between the virtual photon
and an initial quark with momentum fraction $x_B$. The quark is 
knocked off-shell by the virtual photon and becomes on-shell again after 
radiating a gluon. Afterwards the quark or the radiated gluon will have a 
secondary scattering with another soft gluon from the nucleus. 
We denote such a hard-soft process by the diagram in Fig.~\ref{fig7}, 
where the off-shell quark is marked by a filled circle. In the 
double hard processes, on the other hand, 
the quark is on-shell after the first hard scattering with the 
virtual photon. The gluon radiation is then induced by the scattering of 
the quark with another gluon that carries finite momentum fraction $x_L+x_D$.
We denote such hard rescatterings by the diagram in Fig.~\ref{fig8} where an
off-shell parton is again marked by a filled circle.


The conventional categorization of soft and hard rescattering by LQS
is according to the hardness of the virtual partons or the momentum fraction
carried by the second initial parton. The hardness of the off-shell parton 
is characterized by the virtuality that is proportional to $\ell_T^2$.
In the limit $\ell_T\rightarrow 0$, these two processes become identical.
That is why they cancel each other completely as we see from the final 
results  [Eqs.~(\ref{eq:res1}), (\ref{eq:res2}),
(\ref{eq:res3}), (\ref{eq:hc2}) and (\ref{eq:hc3})]. Only gluon
radiation with finite transverse momentum has non-vanishing contribution
to the radiation matrix elements. In the collinear limit
$\ell_T\rightarrow 0$, the original terms ''soft'' and ''hard'' used by LQS
in the large $\ell_T$ processes becomes ambiguous. But we will
keep using them in this paper for historical reasons and for
consistency with the large $\ell_T$ limit. Physically, one should interpret
the soft rescattering as the processes in which the gluon radiation
is induced by the original first hard scattering ($\gamma$-quark), while the
hard rescattering as the processes in which the gluon radiation is induced
by the rescattering, as illustrated in Figs.~\ref{fig7} and \ref{fig8}.
The classification of these two different processes according to the pole 
structure of radiation matrix elements is most convenient in discussing 
the interference effects.

We define the formation time of the gluon radiation as
\begin{equation}
\tau_f \equiv \frac{2q^- z(1-z)}{\ell_T^2}\, . \label{eq:form}
\end{equation}
For very collinear ($\ell_T\rightarrow 0$) gluon radiation
the formation time can become much larger than the relative distance
between the two scattering. Then the two radiation processes should have
destructive interference, leading to the LPM interference effect. One
can see that our results [Eqs.~(\ref{eq:res1}), (\ref{eq:res2}),
(\ref{eq:res3}), (\ref{eq:hc2}) and (\ref{eq:hc3})] have a clear manifest 
of such an LPM interference effect due to the cancellation by the 
interferences between hard-soft and double hard scattering.
Since the gluon's formation time should be smaller than the nuclear
size $MR_A/p^+$ in the given frame, the gluon should then have
a minimum transverse momentum $\ell_T^2\sim Q^2/MR_A$, as we
will demonstrate later in this paper. 

We now reorganize the contributions of different processes according to
the above classification of hard-soft and double hard rescattering. 
We will list our complete calculation of quark-gluon double scattering 
in the following. To help
understand the reorganization of different processes we provide in the
Appendix all the radiation amplitudes induced by single, double and triple
scattering. There we use the approach of helicity amplitudes for high-energy
parton scattering and the results are the same as our complete calculation
in the limit of soft radiation ($z_g=1-z\rightarrow 0$).

All the central-cut diagrams are shown in Fig.~\ref{fig9}. The contributions
from double-hard (Fig.~\ref{fig9}a) and hard-soft scattering 
(Fig.~\ref{fig9}b) are,
\begin{eqnarray}
\overline{H}^D_{CH}&=&\int d\ell_T^2
\frac{\alpha_s}{2\pi} \frac{1+z^2}{1-z}
e^{i(x+x_L)p^+y^-+ix_Dp^+(y_1^- - y_2^-)}\frac{2\pi\alpha_s}{N_c} 
\theta(-y_2^-)\theta(y^- - y_1^-)\nonumber \\
&\times&C_A\frac{k_T^2}{\ell_T^2(\vec{\ell_T}-\vec{k_T})^2}
 e^{-ix_Lp^+(y^- -y_1^- +y_2^-)}\, ,
\label{eq:hhc}
\end{eqnarray}
\begin{eqnarray}
\overline{H}^D_{CS}&=&\int d\ell_T^2 
\frac{\alpha_s}{2\pi} \frac{1+z^2}{1-z}
e^{i(x+x_L)p^+y^-+ix_Dp^+(y_1^- - y_2^-)}\frac{2\pi\alpha_s}{N_c} 
\theta(-y_2^-)\theta(y^- - y_1^-)\nonumber \\
&\times&\left\{ C_F\frac{1}{\ell_T^2}+
C_A\frac{1}{(\vec{\ell_T}-\vec{k_T})^2} e^{ix_Dp^+(y^--y_1^- +y_2^-)/(1-z)} 
\right. \nonumber \\
&-&\left. \frac{C_A}{2}\frac{\vec{\ell_T}\cdot(\vec{\ell_T}-\vec{k_T})}
{\ell_T^2(\vec{\ell_T}-\vec{k_T})^2}
\left[e^{ix_Dp^+y_2^-/(1-z)}+e^{ix_Dp^+(y^- - y_1^-)/(1-z)}\right] \right\} 
\, ,
\label{eq:hsc}
\end{eqnarray}
respectively. As we have discussed earlier, the gluon radiation in the 
double-hard scattering is induced by the 
rescattering of an on-shell quark with another
hard gluon. Therefore the contribution has almost the same form as obtained by
Gunion and Bertsch \cite{GB}(GB) for soft gluon bremsstrahlung induced by 
a single elastic scattering of on-shell quarks, except for the phase 
factors and the splitting function which gives the GB result in the 
soft radiation limit ($z_g=1-z\rightarrow 0$). In the processes 
we consider here, the gluon radiation can also be induced by 
the first hard scattering that produces the quark jet. In these 
so-called hard-soft processes, the final quark or gluon then has a second
scattering with another soft gluon from the nucleus. In Eq.~(\ref{eq:hsc}),
the first term comes from quark rescattering, the second term from 
gluon rescattering and the last two terms come from the interferences 
between the first two processes.

\begin{center}
\begin{minipage}[t]{2.7in}
\begin{figure}
\centerline{\psfig{file=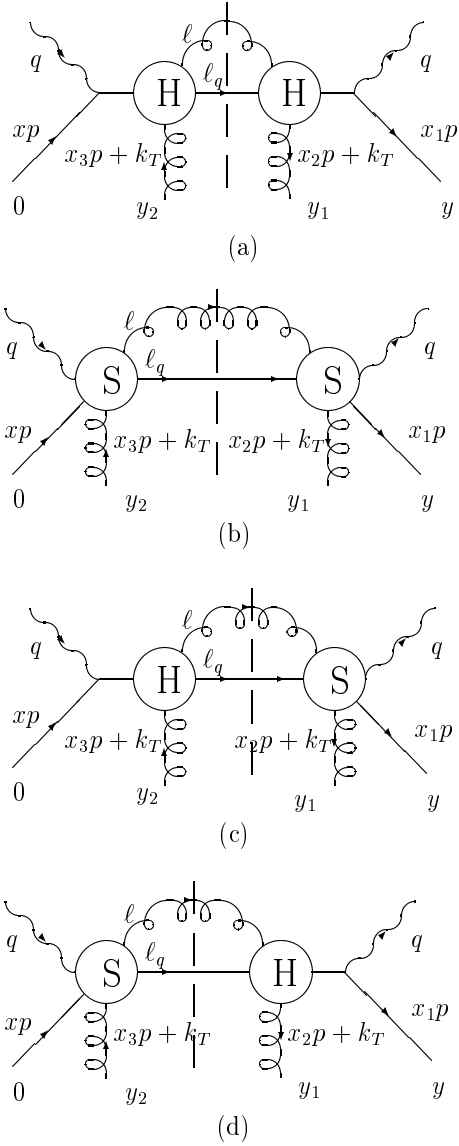,width=2.4in,height=6.0in}}
\caption{Central-cut diagrams for double-hard (a), hard-soft (b) processes 
and their interferences (c) and (d).
The diagramatical representation of the soft and hard rescattering are
shown in Fig.~\ref{fig7} and \ref{fig8}.}
\label{fig9}
\end{figure}
\end{minipage}
\hspace{0.25in}
\begin{minipage}[t]{2.7in}
\begin{figure}
\centerline{\psfig{file=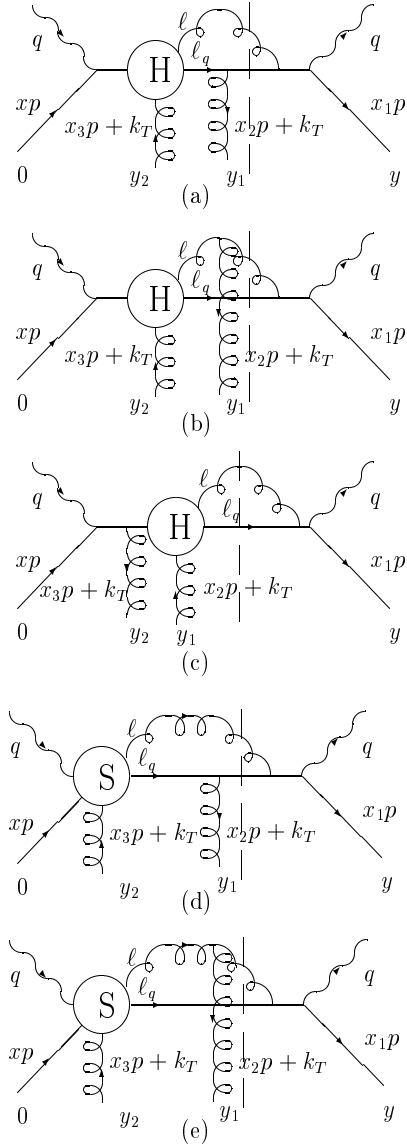,width=2.1in,height=6.0in}}
\caption{Right-cut diagrams that represent interferences between single
and triple scattering. The triple scattering processes involve
double hard and soft (a-c) or hard and double soft (d-e) processes. The 
diagramatical representation of the soft and hard rescattering are
shown in Fig.~\ref{fig7} and \ref{fig8}.}
\label{fig10}
\end{figure}
\end{minipage}
\vspace{0.1in}
\end{center}

There are also two interferences between double hard and hard-soft
scattering as shown in Figs.~\ref{fig9}c and \ref{fig9}d. 
Their contributions are,
\begin{eqnarray}
\overline{H}^D_{CI_2}&=&\int d\ell_T^2 
\frac{\alpha_s}{2\pi} \frac{1+z^2}{1-z}
e^{i(x+x_L)p^+y^-+ix_Dp^+(y_1^- - y_2^-)} \frac{2\pi\alpha_s}{N_c}
\theta(-y_2^-)\theta(y^- - y_1^-)\nonumber \\
&\times&(-)\left\{ \frac{C_A}{2}
\frac{\vec{k_T}\cdot(\vec{k_T}-\vec{\ell_T})}
{\ell_T^2(\vec{\ell_T}-\vec{k_T})^2}
+C_A\frac{\vec{k_T}\cdot\vec{\ell_T}}{\ell_T^2(\vec{\ell_T}-\vec{k_T})^2} 
e^{ix_Dp^+(y^--y_1^-)/(1-z)}\right\} e^{-ix_Lp^+y_2^-}\, .
\label{eq:hic2}
\end{eqnarray}
\begin{eqnarray}
\overline{H}^D_{CI_1}&=&\int d\ell_T^2 
\frac{\alpha_s}{2\pi} \frac{1+z^2}{1-z}
e^{i(x+x_L)p^+y^-+ix_Dp^+(y_1^- - y_2^-)}\frac{2\pi\alpha_s}{N_c} 
\theta(-y_2^-)\theta(y^- - y_1^-)\nonumber \\
&\times(-)&\left\{ \frac{C_A}{2}
\frac{\vec{k_T}\cdot(\vec{k_T}-\vec{\ell_T})}
{\ell_T^2(\vec{\ell_T}-\vec{k_T})^2}
+C_A\frac{\vec{k_T}\cdot\vec{\ell_T}}{\ell_T^2(\vec{\ell_T}-\vec{k_T})^2} 
e^{ix_Dp^+y_2^-/(1-z)}\right\} e^{-ix_Lp^+(y^--y_1^-)}\, ,
\label{eq:hic1}
\end{eqnarray}
It is interesting to note that contributions from the double 
hard processes
or gluon radiation induced by secondary hard scattering and the interferences
with hard-soft processes all vanish in the collinear limit of the secondary
scattering with a gluon ($k_T\rightarrow 0$). What remains is the radiation
spectrum induced by the first hard scattering (photon-quark). 
As we will show later,
combining with interferences between single and triple scattering in the same
collinear limit, it gives the eikonal contribution to the next-to-leading order
correction of the single scattering [Eq.~(\ref{eq:h11})] corresponding to the 
first term in the collinear expansion in Eq.~(\ref{eq-expand1}). For finite
$k_T$ the above radiation spectra exhibit interesting interference patterns
which were discussed in detail by Gyulassy, L\'evai and Vitev \cite{GLV}
and Wiedemann \cite{wied} in the framework of the GW static 
color-screened potential model. In the present paper, we want to
study their contribution to the higher-twist corrections to the fragmentation
functions.

To complete our calculation we also have to consider all the 
interferences between single and triple scattering. 
These correspond to the asymmetrical-cut diagrams.
The right-cut and left-cut dialgrams in Fig.~\ref{fig4} are
just two examples.
After similar reorganization of all different processes, we list in 
Figs.~\ref{fig10} all the possible right-cut interference diagrams. 
The contributions from the first three right-cut diagrams 
in Fig.~\ref{fig10}, which involve double hard scattering, are
\begin{eqnarray}
\overline{H}^D_{RH}&=&\int d\ell_T^2 
\frac{\alpha_s}{2\pi} \frac{1+z^2}{1-z}
e^{i(x+x_L)p^+y^-+ix_Dp^+(y_1^- - y_2^-)}\frac{2\pi\alpha_s}{N_c} 
\theta(-y_2^-)\theta(y_2^- - y_1^-)\nonumber \\
&\times&(-)
\frac{\vec{k_T}\cdot(\vec{k_T}-\vec{\ell_T})}{\ell_T^2(\vec{\ell_T}-\vec{k_T})^2}
\left[\frac{C_A}{2}e^{i(x_D^0-x_D-x_L)p^+(y_1^--y_2^-)}
-\frac{C_A}{2}\right. \nonumber \\
&-&\left. C_A e^{-i(1-z/(1-z))x_Dp^+(y_1^- - y_2^-)} \right]e^{-ix_Lp^+y_2^-}\, ,
\label{eq:hhr}
\end{eqnarray}
where $x_D^0=k_T^2/2p^+q^-$. We have made the variable change 
$k_T\rightarrow -k_T$ in some of the contributions in order to obtain a
more compact form of the final result. The contributions from the two hard-soft
processes in the right-cut diagrams in Fig.~\ref{fig10} are
\begin{eqnarray}
\overline{H}^D_{RS}&=&\int d\ell_T^2
\frac{\alpha_s}{2\pi} \frac{1+z^2}{1-z}
e^{i(x+x_L)p^+y^-+ix_Dp^+(y_1^- - y_2^-)} \frac{2\pi\alpha_s}{N_c}
\theta(-y_2^-)\theta(y_2^- - y_1^-)\nonumber \\
&\times&(-) \left\{
\frac{1}{\ell_T^2}\left[C_F + C_A e^{-i(1-z/(1-z))x_Dp^+(y_1^--y_2^-)}\right]
\right. \nonumber \\
&-&\left.
\frac{\vec{\ell_T}\cdot(\vec{\ell_T}-\vec{k_T})}{\ell_T^2(\vec{\ell_T}-\vec{k_T})^2}
\frac{C_A}{2}e^{ix_Dp^+y_2^-/(1-z)}
\left[1+e^{-i(1-z/(1-z))x_Dp^+(y_1^--y_2^-)}\right]
\right\}\, .
\label{eq:hsr}
\end{eqnarray}
Similarly, contributions from the left-cut diagrams, which are just
the complex conjugates of the right-cut diagrams in Fig.\ref{fig10}, are,
\begin{eqnarray}
\overline{H}^D_{LH}&=&\int d\ell_T^2 
\frac{\alpha_s}{2\pi} \frac{1+z^2}{1-z}
e^{i(x+x_L)p^+y^-+ix_Dp^+(y_1^- - y_2^-)} \frac{2\pi\alpha_s}{N_c}
\theta(y^- -y_1^-)\theta(y_1^- - y_2^-)\nonumber \\
&\times&(-)
\frac{\vec{k_T}\cdot(\vec{k_T}-\vec{\ell_T})}{\ell_T^2(\vec{\ell_T}-\vec{k_T})^2}
\left[\frac{C_A}{2}e^{i(x_D^0-x_D-x_L)p^+(y_1^--y_2^-)}
-\frac{C_A}{2} \right. \nonumber \\
&-&\left. C_A e^{-i(1-z/(1-z))x_Dp^+(y_1^- - y_2^-)} 
\right]e^{-ix_Lp^+(y^--y_1^-)} \, ,
\label{eq:hhl}
\end{eqnarray}
\begin{eqnarray}
\overline{H}^D_{LS}&=&\int d\ell_T^2 
\frac{\alpha_s}{2\pi} \frac{1+z^2}{1-z}
e^{i(x+x_L)p^+y^-+ix_Dp^+(y_1^- - y_2^-)} \frac{2\pi\alpha_s}{N_c}
\theta(y^- -y_1^-)\theta(y_1^- - y_2^-)\nonumber \\
&\times&(-) \left\{
\frac{1}{\ell_T^2}\left[C_F + C_A e^{-i(1-z/(1-z))x_Dp^+(y_1^--y_2^-)}\right]
\right. \nonumber \\
&-&\left.
\frac{\vec{\ell_T}\cdot(\vec{\ell_T}-\vec{k_T})}{\ell_T^2(\vec{\ell_T}-\vec{k_T})^2}
\frac{C_A}{2}e^{ix_Dp^+(y^--y_1^-)/(1-z)}
\left[1+e^{-i(1-z/(1-z))x_Dp^+(y_1^--y_2^-)}\right]
\right\}\, .
\label{eq:hsl}
\end{eqnarray}

\subsection{Collinear Expansion}
Following the procedure of extracting the next-leading-twist contributions
to the semi-inclusive DIS cross section, as outlined in Sec.~\ref{secii},
we now expand the hard part in $k_T$
according to Eq.~(\ref{eq-expand1}). To simplify our notation, we factor
out all the $\theta$-functions and $k_T$-independent part of 
$\overline{H}^D_{C,R,L}$(the sum of all the contributions from central-cut,
right-cut or left-cut diagrams) and define
\begin{eqnarray}
\overline{H}^D&=&\int d\ell_T^2 \frac{\alpha_s}{2\pi} \frac{1+z^2}{1-z}
e^{i(x+x_L)p^+y^- +ix_Dp^+(y_1^- - y_2^-)} \frac{2\pi\alpha_s}{N_c}
\nonumber \\
&\times&\left[H^D_C\theta(-y_2^-)\theta(y^--y_1^-)
-H^D_R\theta(-y_2^-)\theta(y_2^--y_1^-)
-H^D_L\theta(y^--y_1^-)\theta(y_1^--y_2^-)\right]
\, . \label{eq:htheta}
\end{eqnarray}
From Eqs.~(\ref{eq:hhc})-(\ref{eq:hsl}), we have
\begin{equation}
\overline{H}^D_C(k_T=0)=\overline{H}^D_R(k_T=0)=\overline{H}^D_L(k_T=0)=
\frac{C_F}{\ell_T^2} \, .
\end{equation}
They all come from hard-soft processes. As we have pointed out before, 
double-hard scattering processes and their interferences all vanish 
when $k_T=0$. Note that the combination of $\theta$-functions,
\begin{eqnarray}
& &\int dy_1^-dy_2^-\left[\theta(-y_2^-)\theta(y_2^--y_1^-)
+\theta(y^--y_1^-)\theta(y_1^--y_2^-)-\theta(-y_2^-)\theta(y^--y_1^-)
\right] \nonumber \\
&=&\int_0^{y^-}dy_1^-\int_0^{y_1^-}dy_2^- \; , \label{eq:theta}
\end{eqnarray}
is an ordered integral limited by the value of $y^-$. We will refer to
any terms that are proportional to the above combination of $\theta$-functions
as contact contributions.
Combining with Eqs.~(\ref{WD1}) and (\ref{eq:hc0}), the first term in 
the $k_T$ expansion in Eq.~(\ref{eq:htheta}), $\overline{H}^D(k_T=0)$,  
gives the eikonal contribution to the next-leading-order correction of single 
scattering in Eq.~(\ref{eq:h11}). Such eikonal contributions involving 
non-physical gauge fields do not correspond to any physical
double scattering. They only make the final results gauge invariant and can
be gauged away. However, we do see the importance of including all the
interferences (right-cut and left-cut diagrams) between single and triple
scattering. This is essentially a special case of the generalized proof of 
factorization of leading-twist contributions in DIS \cite{factorize}
where contributions from any number of soft gluon rescatterings can be
eikonalized.

The dominant contributions to the double quark-gluon scattering come from 
the quadratic term in the $k_T$ expansion of $\overline{H}^D$ in 
Eq.~(\ref{eq-expand2}). We note that these high-twist terms generally
involve two additional spatial integrations with respect to 
$y_1^-$ and $y_2^-$. If no
restriction is imposed, these integrations will only be limited by the
nuclear size and thus will produce nuclear enhancement as compared to
DIS with a nucleon target.  Because of the rapid oscillation of $e^{ixp^+y^-}$,
any term proportional to Eq.~(\ref{eq:theta}) that limits the integration 
over $y_1^-$ and $y_2^-$ to the value of $y^-$ will not have nuclear 
enhancement. Neglecting any term that is proportional to
Eq.~(\ref{eq:theta}) (contact term) 
and keeping the leading term when $\ell_T\rightarrow 0$, we have
\begin{eqnarray}
\nabla^2_{k_T}H^D_C|_{k_T=0}&=&\frac{4C_A}{\ell_T^4}(1-e^{-ix_Lp^+y_2^-})
(1-e^{-ix_Lp^+(y^--y_1^-)}) +{\cal O}(x_B/Q^2\ell_T^2)\, ,\nonumber  \\
\nabla^2_{k_T}H^D_L|_{k_T=0}&=&0 +{\cal O}(x_B/Q^2\ell_T^2) \, , \nonumber \\
\nabla^2_{k_T}H^D_R|_{k_T=0}&=&0 +{\cal O}(x_B/Q^2\ell_T^2) \, .
\label{eq:hexpand}
\end{eqnarray}
The four terms in $H^D_C$ correspond to hard-soft, double hard scattering
and their interferences.
Substituting the above in Eq.~(\ref{eq:hc0}) and (\ref{eq-expand2}), we
have the leading contribution to the semi-inclusive tensor from double
quark-gluon scattering with quark fragmentation
\begin{eqnarray}
\frac{W_{\mu\nu}^{D,q}}{dz_h}
&=&\sum_q \,\int dx H^{(0)}_{\mu\nu}(xp,q)
\int_{z_h}^1\frac{dz}{z}D_{q\rightarrow h}(z_h/z)
\frac{\alpha_s}{2\pi} C_A\,\frac{1+z^2}{1-z} \nonumber \\
&\times&\int \frac{d\ell_T^2}{\ell_T^4} \frac{2\pi\alpha_s}{N_c} 
T^A_{qg}(x,x_L) +({\rm virtual\,\, correction})\, , \label{eq:WD-fq}
\end{eqnarray}
where
\begin{eqnarray}
T^A_{qg}(x,x_L)&=& \int \frac{dy^{-}}{2\pi}\, dy_1^-dy_2^-
e^{i(x+x_L)p^+y^-}(1-e^{-ix_Lp^+y_2^-})
(1-e^{-ix_Lp^+(y^--y_1^-)}) \nonumber \\
&\frac{1}{2}&\langle A | \bar{\psi}_q(0)\,
\gamma^+\, F_{\sigma}^{\ +}(y_{2}^{-})\, F^{+\sigma}(y_1^{-})\,\psi_q(y^{-})
| A\rangle \theta(-y_2^-)\theta(y^- -y_1^-)
\label{eq:qgmatrix}
\end{eqnarray}
is the quark-gluon correlation function which 
essentially contains four independent
twist-four parton matrix elements in a nucleus. Since $x_L$ as defined in
Eq.~(\ref{eq:xld}) depends on $\ell_T$ and $z$, the matrix element 
$T^A_{qg}(x,x_L)$ has an implicit dependence on $z$ and $\ell_T$.
The contribution from gluon fragmentation is similarly
\begin{eqnarray}
\frac{W_{\mu\nu}^{D,g}}{dz_h}
&=&\sum_q \,\int dx H^{(0)}_{\mu\nu}(x,p,q)
\int_{z_h}^1\frac{dz}{z}D_{g\rightarrow h}(z_h/z)
\frac{\alpha_s}{2\pi} C_A\,\frac{1+(1-z)^2}{z} \nonumber \\
&\times&\int \frac{d\ell_T^2}{\ell_T^4} \frac{2\pi\alpha_s}{N_c} T^A_{qg}(x,x_L) 
+({\rm virtual\,\, correction}) \, . \label{eq:WD-fg}
\end{eqnarray}

The above contribution from
double quark-gluon scattering is very similar to the next-to-leading order
single scattering in Eq.~(\ref{eq:h11}). Even the splitting function has the
same form except for the color factor. 
The contribution is, however, proportional to 
the twist-four parton matrix elements. Since there is not much restriction on
the spatial integral, such twist-four matrix elements will give the nuclear
enhancement we are looking for. We will return later for more discussions on
the $A$ dependence of the double scattering contribution.

In the process of a collinear expansion, we have kept $\ell_T$ finite and
took the limit $k_T\rightarrow 0$. As a consequence, the gluon field
in one of the twist-four parton matrix elements in Eq.~(\ref{eq:qgmatrix})
carries zero momentum, corresponding to the soft-hard process. Later in
this paper we will discuss how to estimate these twist-four matrix
elements. Assuming a factorized form, they will be proportional to the
product of quark and gluon distributions in a nucleon. However, the
gluon distribution $xf_g(x)$ at $x=0$ is not defined in QCD. This is
because we dropped higher order terms of the collinear expansion in 
Eq.~(\ref{eq:hexpand}). These terms are considered higher-twist
contributions in the LQS framework. As a remedy to the problem, we
simply resum a subset of the higher-twist terms in the collinear
expansion and restore the phase factors such as $\exp(ix_Tp^+y^-)$ 
which is related to the intrinsic transverse momentum of the initial
partons, $x_T\equiv \langle k_T^2\rangle/2p^+q^-z$. In this case,
soft gluon fields in the parton matrix elements will carry a
fractional momentum $x_T$. Similarly, one can also restore the
collinear structure of the scattering amplitude by replacing $1/\ell_T^4$
in Eq.~(\ref{eq:hexpand}) with $1/\ell_T^2(\ell_T^2+\langle k_T^2\rangle)$.

\begin{figure}
\centerline{\psfig{file=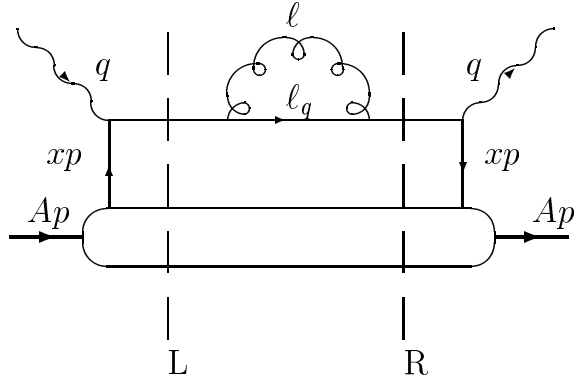,width=3in,height=2.0in}}
\caption{Diagrams for the virtual correction to single scattering with two
possible cuts.}
\label{fig:virtual}
\end{figure}

\subsection{Virtual Corrections}
So far we have not considered virtual corrections which will ensure the
final result to be infrared safe. The calculation of the virtual correction
to the single scattering in Fig.~\ref{fig:virtual} is very similar to
the real correction in Fig.~\ref{fig3}. One can easily find,
\begin{equation}
\frac{dW^{S(v)}_{\mu\nu}}{dz_h}
=-\sum_q \int dx f_q^A(x) H^{(0)}_{\mu\nu}(x,p,q) 
D_{q\rightarrow h}(z_h) \int_0^{\mu^2} 
\frac{d\ell_T^2}{\ell_T^2} \frac{\alpha_s}{2\pi} \int_{0}^1dz\,
C_F \frac{1+z^2}{1-z} \, . \label{eq:vs}
\end{equation}
One can see that the integral over $z$ is infrared divergent at
$z=1$ or $z_g=0$. However, such divergency cancels exactly the 
infrared divergency in Eq.~(\ref{eq:h11}) of the radiative contribution
to the single scattering process. The sum is then infrared safe. Using
the definition of the $+$functions \cite{dglap},
\begin{equation}
\int_0^1 dz \frac{F(z)}{(1-z)_+} \equiv \int_0^1 dz \frac{F(z)-F(1)}{1-z}
\label{eq:plus}
\end{equation}
with $F(z)$ being any function which is sufficiently smooth at $z=1$,
one can rewrite \cite{JCJQ} the sum of radiative and virtual correction
[Eq.~(\ref{eq:s-sum})] in the following form
\begin{eqnarray}
D_{q\rightarrow h}(z_h,\mu^2)&=&D_{q\rightarrow h}(z_h)+
\int_0^{\mu^2} \frac{d\ell_T^2}{\ell_T^2} 
\frac{\alpha_s}{2\pi} \int_{z_h}^1 \frac{dz}{z}
\left[ \gamma_{q\rightarrow qg}(z) D_{q\rightarrow h}(z_h/z)\right. 
\nonumber \\
&+& \left.  \gamma_{q\rightarrow gq}(z)D_{g\rightarrow h}(z_h/z)\right] \, .
\label{eq:s-sum2}
\end{eqnarray}
The splitting functions are defined as
\begin{eqnarray}
\gamma_{q\rightarrow qg}(z) &=& C_F\left[\frac{1+z^2}{(1-z)_+} 
  + \frac{3}{2}\delta(1-z)\right] \; , \label{eq:split1}\\
\gamma_{q\rightarrow gq}(z) &=& \gamma_{q\rightarrow qg}(1-z) \; .
 \label{eq:split2}
\end{eqnarray}
Therefore, with the definition of the $+$function, the 
$\delta$-function terms in the splitting functions take into account 
the self-energy virtual correction which cancels the infrared divergences 
from the radiative processes. The final renormalized quark fragmentation
$D_{q\rightarrow h}(z_h,\mu^2)$ satisfies the DGLAP evolution equations
in Eq.~(\ref{eq:ap1}).

When cast into the DGLAP evolution equation in Eq.~(\ref{eq:ap1}), 
the splitting function from the real correction can be interpreted as
the probability for the quark to radiate a gluon with momentum fraction
$1-z$. Then one must also take into account the probability of no
gluon radiation in the evolution to ensure unitarity. Such unitarity
requirement gives rise to the same virtual correction as calculated from
the diagram in Fig.~\ref{fig:virtual}. In the double scattering case, we
will use the same unitarity requirement to obtain virtual corrections.
The virtual contribution to the quark fragmentation in double
scattering processes is, for example,
\begin{equation}
\frac{W_{\mu\nu}^{D(v),q}}{dz_h}
=-\sum_q \,\int dx H^{(0)}_{\mu\nu}(xp,q)
D_{q\rightarrow h}(z_h) \frac{\alpha_s}{2\pi} C_A\int_{0}^1dz\,\frac{1+z^2}{1-z}
\int \frac{d\ell_T^2}{\ell_T^4} \frac{2\pi\alpha_s}{N_c} T^A_{qg}(x,x_L) 
\, . \label{eq:WD-fqv1}
\end{equation}
One can single out the infrared divergent part by rewriting the integral
\begin{eqnarray}
\int_{0}^1dz\,\frac{1+z^2}{1-z}T^A_{qg}(x,x_L)
&=&T^A_{qg}(x,x_L)|_{z=1}\int_{0}^1dz\frac{2}{1-z}
-\Delta T^A_{qg}(x,\ell_T^2) \; , \nonumber \\
\Delta T^A_{qg}(x,\ell_T^2) &\equiv &
\int_0^1 dz\frac{1}{1-z}\left[ 2 T^A_{qg}(x,x_L)|_{z=1}
-(1+z^2) T^A_{qg}(x,x_L)\right] \, . \label{eq:vsplit}
\end{eqnarray}
The second term is finite since $T^A_{qg}(x,x_L)$ is a smooth function of $z$.
The first term can be combined with the radiative contribution 
in Eq.~(\ref{eq:WD-fq}) to cancel the infrared divergency. 
With the help of the $+$function, the final result can be expressed as 
\begin{eqnarray}
\frac{W_{\mu\nu}^{D,q}}{dz_h}
&=&\sum_q \,\int dx H^{(0)}_{\mu\nu}(xp,q)
\frac{2\pi\alpha_s}{N_c}\int \frac{d\ell_T^2}{\ell_T^4}  
\int_{z_h}^1\frac{dz}{z}D_{q\rightarrow h}(z_h/z) \nonumber \\
&\times& \frac{\alpha_s}{2\pi} C_A\left[
\frac{1+z^2}{(1-z)_+}T^A_{qg}(x,x_L)+\delta(z-1)
\Delta T^A_{qg}(x,\ell_T^2) \right] \, .\label{eq:WD-fq2}
\end{eqnarray}
Here the implicit $z$-dependence of $T^A_{qg}(x,x_L)$ 
plays an important role in the
final result. The above integrand will be proportional
to the splitting function for single scattering in Eq.~(\ref{eq:split1}) if
one ignores the $z$ dependence of $T^A_{qg}(x,x_L)$.
Similarly, the final result for contributions from gluon fragmentation is
\begin{eqnarray}
\frac{W_{\mu\nu}^{D,g}}{dz_h}
&=&\sum_q \,\int dx H^{(0)}_{\mu\nu}(xp,q)
\frac{2\pi\alpha_s}{N_c}\int \frac{d\ell_T^2}{\ell_T^4}  
\int_{z_h}^1\frac{dz}{z}D_{g\rightarrow h}(z_h/z) \nonumber \\
&\times& \frac{\alpha_s}{2\pi} C_A\left[
\frac{1+(1-z)^2}{z_+}T^A_{qg}(x,x_L)+\delta(z)
\Delta T^A_{qg}(x,\ell_T^2) \right] \, , \label{eq:WD-fg2}
\end{eqnarray}
where we have used the fact that $x_L$ in Eq.~(\ref{eq:xld}) is invariant
under the transform $z\rightarrow 1-z$ and so is $T^A_{qg}(x,x_L)$.

\begin{figure}
\centerline{\psfig{file=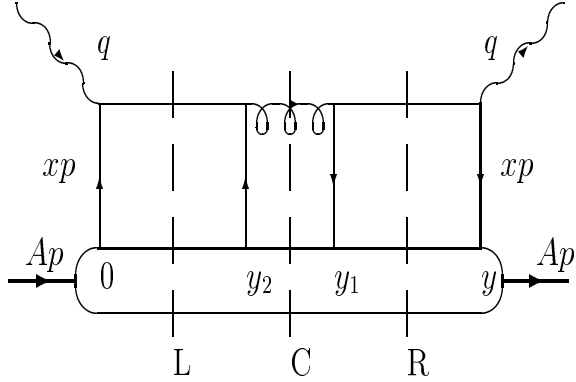,width=3in,height=2.0in}}
\caption{The diagrams for leading order quark-quark double scattering.}
\label{fig:qq0}
\end{figure}

\section{Quark-quark Double Scattering}
We have so far only considered quark-gluon double scattering in a nucleus.
After the first photon-quark hard scattering, the leading quark can also 
rescatter with another quark from the nucleus. 
Such quark-quark double scattering processes also contribute to the 
semi-inclusive DIS at twist-four. Unlike quark-gluon rescattering, quark-quark
double scattering even contributes at the lowest order without gluon radiation.
At the lowest order there is only one kind of quark-quark double scattering
diagram, as shown in Fig.~\ref{fig:qq0} and its crossing variations. 
One can easily calculate their
contributions and obtain
\begin{eqnarray}
\frac{dW^{D(0)}_{qq\mu\nu}}{dz_h}&=&\sum_q \int dx
\int\frac{dy^-}{2\pi}dy_1^-dy_2^- e^{ixp^+y^-}
\langle A|\bar{\psi}_q(0)\frac{\gamma^+}{2}\psi_q(y^-) 
\bar{\psi}_q(y_1^-)\frac{\gamma^+}{2}\psi_q(y_2^-)|A\rangle \nonumber \\
&\times&\frac{2\pi\alpha_s}{N_c} 8C_F\frac{x_B}{Q^2} H^{(0)}_{\mu\nu}(x,p,q)
\left\{D_{g\rightarrow h}(z_h)\theta(-y_2^-)\theta(y^--y_1^-)\right.
 \nonumber \\
&-&\left. D_{q\rightarrow h}(z_h)\left[\theta(y^--y_1^-)\theta(y_1^--y_2^-)
+\theta(-y_2^-)\theta(y_2^--y_1^-)\right] \right\} \, ,
\end{eqnarray}
where to combine all the crossing diagram we use the fact that fields on the
light-cone commute with each other \cite{QS}. 
In the above contributions, the term
proportional to the gluon fragmentation function is from the central-cut
diagrams while the terms containing quark fragmentation function are from
left and right-cut diagrams (or interferences). Using Eq.~(\ref{eq:theta})
and neglecting the contact term, we have
\begin{equation}
\frac{dW^{D(0)}_{qq\mu\nu}}{dz_h}=\sum_q \int dx T^{A(I)}_{qq}(x)
\frac{2\pi\alpha_s}{N_c} 8C_F\frac{x_B}{Q^2} H^{(0)}_{\mu\nu}(x,p,q)
\left[D_{g\rightarrow h}(z_h)-D_{q\rightarrow h}(z_h) \right] \, ,
\label{eq:qq0}
\end{equation}
where
\begin{equation}
T^{A(I)}_{qq}(x)=\int\frac{dy^-}{2\pi}dy_1^-dy_2^- e^{ixp^+y^-}
\langle A|\bar{\psi}_q(0)\frac{\gamma^+}{2}\psi_q(y^-) 
\bar{\psi}_q(y_1^-)\frac{\gamma^+}{2}\psi_q(y_2^-)|A\rangle
\theta(-y_2^-)\theta(y^--y_1^-)
\label{eq:eq:qqmatrix}
\end{equation}
is a four-quark matrix element in a nucleus. This twist-four
contribution is explicitly suppressed by $1/Q^2$ as compared to the
single scattering case. Such leading-order and high-twist 
contributions from quark-quark double scattering essentially mix
the quark and gluon fragmentation functions. In this case, there
is no induced radiation and thus no energy loss. 
However, these contributions will 
change the final differential cross section or semi-inclusive spectrum, 
since quark and gluon fragmentation functions are different.
One thus can consider the case as modification of fragmentation
functions without energy loss.
Furthermore, these processes do not change the integrated total 
cross section. One can verify this by using the momentum sum rule
$\sum_h\int dz_h z_h D_{q,g\rightarrow h}(z_h)=1$. 
This is consistent with a general theorem that final state 
interaction will not change the total cross section.

\begin{figure}
\vspace{0.2in}
\centerline{\psfig{file=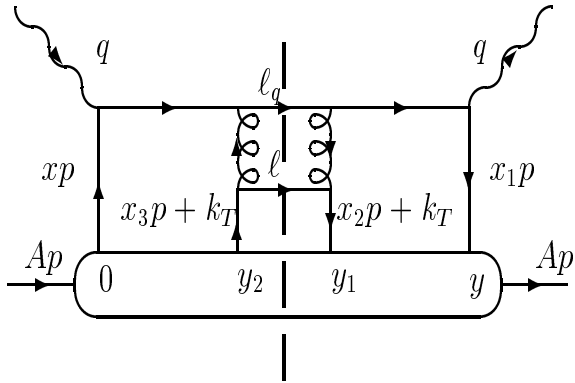,width=3in,height=2.0in}}
\caption{An example diagram for next-to-leading order quark-quark
double scattering.}
\label{fig:qq1}
\vspace{0.2in}
\end{figure}

We also note that contributions from quark-quark double scattering 
are free of any divergences, especially collinear divergency. Therefore,
they will not contribute to the QCD evolution of the effective
parton fragmentation functions in the nuclear medium.

To the next order we have to consider radiative corrections to
the processes in Fig.~\ref{fig:qq0}. In addition we also have to consider
the diagram in Fig.~\ref{fig:qq1}. Again, calculations of these diagrams
are tedious but straightforward. Take the diagram in Fig.~\ref{fig:qq1} for
example: The pole structure is exactly the same as the central-cut diagram
of quark-gluon double scattering in Fig.~\ref{fig4}. 
The resultant contribution is
\begin{eqnarray}
\frac{dW^{D(1)}_{Fig.\ref{fig:qq1}\mu\nu}}{dz_h}&=&\sum_q 
\int_{z_h}^1\frac{dz}{z} D_{q\rightarrow h}(z_h/z)
\int\frac{d\ell_T^2}{\ell_T^2} \frac{\alpha_s}{2\pi}
C_F \frac{1+z^2}{(1-z)^2} \frac{x_B}{Q^2} \nonumber \\
& \times &\frac{2\pi\alpha_s}{N_C} \int dx T^{A(II)}_{qq}(x,x_L)
H^{(0)}_{\mu\nu}(x,p,q)\, ,
\end{eqnarray}
where
\begin{eqnarray}
T^{A(II)}_{qq}(x,x_L)&=&\sum_{q_i}
\int\frac{dy^-}{2\pi}dy_1^-dy_2^- e^{i(x+x_L)p^+y^-}
(1-e^{-ix_Lp^+y_2^-}) (1-e^{-ix_Lp^+(y^--y_1^-)}) \nonumber \\
&\times&\langle A|\bar{\psi}_q(0)\frac{\gamma^+}{2}\psi_q(y^-) 
\bar{\psi}_{q_i}(y_1^-)\frac{\gamma^+}{2}\psi_{q_i}(y_2^-)|A\rangle
\theta(-y_2^-)\theta(y^--y_1^-) \, .
\end{eqnarray}
The four terms with different phase factors in the above equation
correspond to hard-soft, double hard and the interferences. Notice that
the two diagonal terms are defined exactly the same as $T^{A(I)}_{qq}(x)$.
This structure
is exactly the same as in the quark-gluon double scattering. However,
this contribution is only proportional to $1/\ell_T^2$ versus $1/\ell_T^4$
in the quark-gluon double scattering. It is thus suppressed
by $1/Q^2$. This remains to be the case for all quark-quark double 
scattering processes. Since we are
interested only in the collinear behavior of the gluon radiation processes
in order to study the QCD evolution, we will neglect in this paper
all radiative contributions from quark-quark double scattering.
Similarly, contributions proportional to $1/\ell_T^2$ 
from gluon-gluon double scattering can also
be neglected to this approximation, {\it e.g.}, in the collinear
expansion in Eq.~(\ref{eq:hexpand}).

\section{Modified Quark Fragmentation Function}

For a complete result of the semi-inclusive cross section of DIS off a 
nucleus, one should also include the higher-twist contribution to the
quark distributions and their QCD evolution as studied by Mueller and
Qiu \cite{MQ}. Here we will simply replace the leading-twist
quark distributions $f_q^A(x)$ by $\widetilde{f}_q^A(x,\mu_I^2)$ which
contains nuclear modification to the quark distributions and their
QCD evolution and $\mu_I^2$ is the factorization scale for the
quark distributions in a nucleus. We should note that even the 
leading-twist quark distributions $f^A_q(x)$ in a nucleus are different from
$A$ free nucleons [$f^N_q(x)$].

\subsection{Modified Evolution Equations}

Including the higher-twist contributions to the
quark distributions and summing up all the leading contributions from 
single and double scattering processes in 
Eqs.~(\ref{eq:s-sum2}), (\ref{eq:WD-fq2}) and (\ref{eq:WD-fg2}),
we have
\begin{equation}
\frac{dW_{\mu\nu}}{dz_h}=\sum_q \int dx \widetilde{f}_q^A(x,\mu_I^2) 
H^{(0)}_{\mu\nu}(x,p,q)
\widetilde{D}_{q\rightarrow h}(z_h,\mu^2) \label{eq:Wtot}
\end{equation}
as the total semi-inclusive tensor in DIS off a nucleus up to twist-four
corrections.
We define the modified effective quark fragmentation function  as
\begin{eqnarray}
\widetilde{D}_{q\rightarrow h}(z_h,\mu^2)&\equiv& 
D_{q\rightarrow h}(z_h,\mu^2)
+\int_0^{\mu^2} \frac{d\ell_T^2}{\ell_T^2} 
\frac{\alpha_s}{2\pi} \int_{z_h}^1 \frac{dz}{z}
\left[ \Delta\gamma_{q\rightarrow qg}(z,x,x_L,\ell_T^2) 
D_{q\rightarrow h}(z_h/z) \right. \nonumber \\
&+& \left. \Delta\gamma_{q\rightarrow gq}(z,x,x_L,\ell_T^2)
D_{g\rightarrow h}(z_h/z)\right] \, , \label{eq:dmod}
\end{eqnarray}
where $D_{q\rightarrow h}(z_h,\mu^2)$ is given in Eq.~(\ref{eq:s-sum2}) for
leading-twist contributions and
\begin{eqnarray}
\Delta\gamma_{q\rightarrow qg}(z,x,x_L,\ell_T^2)&=&
\left[\frac{1+z^2}{(1-z)_+}T^A_{qg}(x,x_L) + 
\delta(1-z)\Delta T^A_{qg}(x,\ell_T^2) \right]
\frac{C_A2\pi\alpha_s}
{(\ell_T^2+\langle k_T^2\rangle)N_c\widetilde{f}_q^A(x,\mu_I^2)}
\label{eq:dsplit1}\\
\Delta\gamma_{q\rightarrow gq}(z,x,x_L,\ell_T^2) 
&=& \Delta\gamma_{q\rightarrow qg}(1-z,x,x_L,\ell_T^2) \label{eq:dsplit2}.
\end{eqnarray}
The twist-four matrix element $T^A_{qg}(x,x_L)$ is given in 
Eq.~(\ref{eq:qgmatrix}) and $\Delta T^A_{qg}(x,\ell_T^2)$ is 
given in Eq.~(\ref{eq:vsplit}). Since the two-parton correlation 
function $T^A_{qg}$ can involve partons from two different nucleons 
inside the nucleus, it should have a nuclear enhancement that is 
proportional to the nuclear size $R_A$, or 
$T^A_{qg}/\widetilde{f}_q^A \propto A^{1/3}$, as we will show in the next
subsection. Because of the LPM interference structure contained in
$T^A_{qg}(x,x_L)=0$ ($x_L\sim \ell_T^2 \rightarrow 0$), only those
radiated gluons induced by multiple parton scattering
that have finite transverse momentum, $\ell_T^2\sim Q^2/A^{1/3}$ 
contribute to the fragmentation function. Therefore, the nuclear
correction to the fragmentation function in Eq.~\ref{eq:dmod} is
proportional to $\alpha_s A^{2/3}/Q^2$ from double parton scattering.
For large $A$ and $Q^2$, this is the leading nuclear correction.
Since $T_{qg}(x,x_L)/\widetilde{f}_q^A(x)$ is proprotional
to the gluon density per unit transverse area in the nucleus
$\rho_g$ and $\alpha_s/Q^2$ can be considered as the parton rescattering
cross section $\sigma_g$, the leading nuclear correction to the
fragmentation function here is then proportional to $\sigma_g\rho_g R_A$.
This is exactly the same as the expansion parameter in the opacity expansion
approach in Refs.~\cite{GLV,wied}.

We should emphasize here that the factorized form of the semi-inclusive
tensor in Eq.(\ref{eq:Wtot}) only serves to define the effective quark
fragmentation function $\widetilde{D}_{q\rightarrow h}(z_h,\mu^2)$.
Such a modified fragmentation function for final state
particle production has an explicit dependence on the initial parton
distribution through the high-twist double scattering processes. Therefore,
the factorization for semi-inclusive processes in DIS is broken explicitly
at twist-four correction. This is a natural consequence of the 
non-vanishing parton energy loss at twist-four when the leading quark
suffers multiple scattering through the nuclear medium.

Taking the
derivative with respect to the collinear factorization scale $\mu^2$,
we obtain the modified DGLAP evolution equation in leading order of 
$\alpha_s$ for the modified quark fragmentation function,
\begin{eqnarray}
  \frac{\partial \widetilde{D}_{q\rightarrow h}(z_h,\mu^2)}
  {\partial \ln \mu^2}  &=& 
  \frac{\alpha_s}{2\pi} \int^1_{z_h} \frac{dz}{z} 
\left[ \widetilde{\gamma}_{q\rightarrow qg}(z,x,x_L,\mu^2)
\widetilde{D}_{q\rightarrow h}(z_h/z,\mu^2) \right. \nonumber \\
&+&\left.\widetilde{\gamma}_{q\rightarrow gq}(z,x,x_L,\mu^2) 
D_{g\rightarrow h}(z_h/z,\mu^2)\right] \, .  \label{eq:e-ap1}
\end{eqnarray}
The modified splitting functions are defined as
\begin{eqnarray}
\widetilde{\gamma}_{q\rightarrow qg}(z,x,x_L,\mu^2) &=&
\gamma_{q\rightarrow qg}(z)+\Delta\gamma_{q\rightarrow qg}(z,x,x_L,\mu^2)
\label{eq:e-split1}\\
\widetilde{\gamma}_{g\rightarrow gq}(z,x,x_L,\mu^2) 
&=& \widetilde{\gamma}_{q\rightarrow qg}(1-z,x,x_L,\mu^2) 
\label{eq:e-split2},
\end{eqnarray}
where $\gamma_{q\rightarrow qg}(z)$ given in Eq.~(\ref{eq:split1}) is the
splitting function in single scattering processes.
If one only considers single-jet events in DIS, 
the gluon fragmentation function
does not contribute in the leading order. 
Before we carry out similar calculations for the higher-twist correction
to the DGLAP evolution of the gluon fragmentation functions,
we may assume that the gluon
fragmentation function which enters into the above equation follows the 
normal DGLAP evolution,
\begin{equation}
\frac{\partial D_{g\rightarrow h}(z_h,\mu^2)}{\partial \ln \mu^2}  = 
\frac{\alpha_s}{2\pi} \int^1_{z_h} \frac{dz}{z} \left[
    \sum_{q=1}^{2n_f} \gamma_{g\rightarrow q\bar{q}}(z)
  \widetilde{D}_{q\rightarrow h}(z_h/z,\mu^2) + \gamma_{g\rightarrow gg}(z)
 D_{g\rightarrow h}(z_h/z,\mu^2)\right] \label{eq:e-ap2},
\end{equation}
where the normal splitting functions are
\begin{eqnarray}
\gamma_{g\rightarrow q\bar{q}}&=& \frac{1}{2}[z^2+(1-z)^2]\, ,\\
\gamma_{g\rightarrow gg}&=&2C_A\left[ \frac{z}{(1-z)_+} +\frac{1-z}{z}+z(1-z)
\frac{1}{12}(11-\frac{2}{3}n_f)\delta(z-1)\right]\, ,
\end{eqnarray}
and $n_f$ is the number of quark flavors. All the fragmentation functions
obey the momentum sum rule,
\begin{equation}
  \int_0^1 dz \sum_h z\,D_{a\rightarrow h}(z,\mu^2)=1 \, .
\end{equation}
One can check that the modified quark fragmentation function
$\widetilde{D}_{q\rightarrow h}(z,\mu^2)$ 
in Eq.~(\ref{eq:dmod}) still satisfies the
momentum sum rule. Such a momentum sum rule might seem count-intuitive since
there is momentum transfer of $x_Lp^+$ to the fragmentating quark
from the second partons in the nucleus in the double-hard processes. 
However, one should note that
the initial quark from the nucleus carries momentum $x_Bp^+$ in this case.
In single scattering and hard-soft processes, however, the initial quark 
carries momentum $(x_B+x_L)p^+$. So the total momentum transfer from the
nucleus to the quark is the same in both single and double scattering.
The momentum sum rule for the modified fragmentation functions should
still be valid. 

Compared to Eq.~(\ref{eq:ap1}), the renormalization equation for the 
modified quark fragmentation function in Eq.~(\ref{eq:e-ap1}) is similar to 
the original DGLAP evolution equation for the fragmentation functions in 
vacuum. However, the modified splitting functions $\widetilde{\gamma}$
[Eq.~(\ref{eq:e-split1})] have an extra term $\Delta\gamma$ from induced
gluon radiation. Similarly to the ordinary parton cascade in vacuum, the
induced radiation will soften the modified quark fragmentation function.
As we have argued in the Introduction, such softening will be the only 
experimentally measurable consequence of quark energy loss in a medium.
Another important difference between the modified and the original DGLAP
evolution equations is that the induced splitting function $\Delta\gamma$
depends on the twist-four parton matrix elements $T^A_{qg}(x,x_L)$
of the nucleus.
Because of the interferences between hard-soft and double-hard processes,
$T^A_{qg}(x,x_L)$ explicitly manifests the LPM effect which modifies the
gluon radiation spectra such that the induced splitting functions are
also modified from their form in vacuum. The induced splitting functions
also explicitly depend on the factorization scale $\mu$.

\subsection{Twist-four Parton Matrix Elements}

As we have observed, modified DGLAP evolution equations for parton 
fragmentation functions depend on both the parton distribution 
$\widetilde{f}_q^A(x,\mu_I^2)$ and two-parton correlation
functions $T^A_{qg}(x,x_L)$.  Naturally, they are essential to
any numerical solution to the modified DGLAP evolution equations for the
fragmentation functions. One also has to know them, especially their
dependence on the nuclear size, in order to have any numerical estimate
of the effective quark energy loss.

As defined in Eq.~(\ref{eq:qgmatrix}), the parton correlation function 
$T^A_{qg}(x,x_L)$ essentially contains four independent twist-four parton
matrix elements in a nucleus. They are in principle not calculable and
can only be measured independently in experiments just like parton 
distributions. However, under certain assumptions, one might be able
to relate the parton correlation functions to single parton distributions
in the nuclei and in the meantime obtain the $A$-dependence of the 
parton correlations. There are four parton field operators and three spatial
(in the longitudinal direction) integrations in $T^A_{qg}(x,x_L)$. We
assume that the nuclear wave function can be expressed as a multiple-nucleon
state and each nucleon is in a color singlet state. In this case, there
should not be long range color correlation between two nucleons due to color
confinement. In other words, the interaction between nucleons inside a 
nucleus can only be mediated via color-singlet objects. Consider first 
the case where the quark and gluon fields operate on different nucleons 
inside the nucleus. This means that the double scattering happens with two
partons from two different nucleons. Because of the color confinement, 
$y_1$ and $y_2$ 
in Eq.~(\ref{eq:qgmatrix}) are restricted such that $|y_1-y_2|\leq r_N$, 
where $r_N$ is the radius of a nucleon. The integrations over $y_1$ and $y_2$ 
should, therefore, give the length scale of $r_NR_A$, where 
$R_A=1.12A^{1/3}$ is the radius of the nucleus. In this case the
correlation functions in Eq.~(\ref{eq:qgmatrix}) should be 
approximately proportional to $A^{4/3}$. If both the quark and gluon 
fields operate on the same nucleon state inside the nucleus, $y_1$ 
and $y_2$ are both restricted to be within the size of the nucleon. 
Then the correlation functions should only be roughly proportional to $A$.
This corresponds to double scattering with two partons in the same nucleon.
As we have discussed before, we will neglect these contributions because
they yield no nuclear enhancement.

Among the four independent matrix elements in $T^A_{qg}(x,x_L)$, there
are two diagonal terms, one proportional to a phase factor 
$\exp[i(x+x_L)p^+y^- + ix_Tp^+(y_1^- - y_2^-)]$ and another to
$\exp[ixp^+y^- + i(x_L+x_T)p^+(y_1^- - y_2^-)]$. The first term corresponds
to hard-soft double scattering processes in which the gluon carrying 
momentum fraction $x_T$ is soft. The second term comes from double-hard
scattering in which the gluon carries large momentum fraction $x_L+x_T$.
In this paper, we adopt the approximation used in Ref.~\cite{LQS}. Under
such an approximation, one simply relates the two-parton correlations to
the product of two single parton distributions,
\begin{eqnarray}
\int \frac{dy^{-}}{2\pi}\, dy_1^-dy_2^- 
&&e^{ix_1p^+y^-+ix_2p^+(y_1^--y_2^-)} \nonumber \\
&\times& \frac{1}{2}\langle A | \bar{\psi}_q(0)\,
\gamma^+\, F_{\sigma}^{\ +}(y_{2}^{-})\, F^{+\sigma}(y_1^{-})\,\psi_q(y^{-})
| A\rangle \theta(-y_2^-)\theta(y^--y_1^-) \nonumber \\
&=&\frac{C}{x_A} f_q^A(x_1)\, x_2f_g^N(x_2)\, , \label{eq:t4matrix}
\end{eqnarray}
where $x_A=1/MR_A$, $f_q^A(x)$ is the quark distribution inside a nucleus 
as defined in Eq.~(\ref{eq:fquark}), and $f_g^N(x)$ is the gluon distribution
inside a nucleon as defined by
\begin{equation}
f_g^N(x)\equiv \frac{1}{xp^+}\int \frac{dy^-}{2\pi} e^{ixp^+y^-}
\langle N|F_{\sigma}^{\ +}(0)\, F^{ +\sigma}(y^{-})|N\rangle\, .
\end{equation}
$C$ is assumed to be a constant, reflecting the strength of two-parton 
correlation inside a nucleus. As we have argued, $C$ should be proportional
to the nucleon radius $r_N$, characterizing the confinement scale.
We also assume that the nuclear distribution takes a Gaussian form 
$\rho(r) \sim \exp(-r^2/2R_A^2)$. In terms of the light-cone coordinates,
we have then $\rho(y^-)=\rho_0 \exp({y^-}^2/2{R^-_A}^2)$  
where $R^-_A=\sqrt{2}R_AM/p^+$ and $M$ is the nucleon mass. The
distribution is normalized such that
\begin{equation}
\int_0^\infty dy^- \rho_0 e^{-{y^-}^2/2{R_A^-}^2}=1.
\end{equation}
With such a distribution, the third integration in Eq.~(\ref{eq:t4matrix}) 
gives $\int dy_2^-\sim R_A^-\sim 1/x_Ap^+$.

In addition to the two diagonal terms, there are also two off-diagonal
matrix elements in $T^A_{qg}(x,x_L)$ which come from the interferences
between hard-soft and double-hard scattering processes. These two
new matrix elements introduced by the interferences have never been
studied before in DIS. Similar matrix elements have been discussed before
in the context of $pA$ collisions \cite{KSM89}. For finite value of $x_L$,
this involves off-diagonal parton matrix elements and they
should be related to the so-called skewed parton distributions of
nucleons \cite{skew}. For a rough estimate here, we can generalize 
the approximation on parton correlations
in Ref.~\cite{LQS} to the off-diagonal matrix elements. As we have pointed
out in the calculation of the cut-diagrams in Fig.~\ref{fig4}, in the
interferences between double-hard and hard-soft processes, there is actually
momentum flow of $x_Lp^+$ between the two nucleons where 
the initial quark and gluon come from. 
Without strong long range two-nucleon correlation inside a
nucleus, the amount of momentum flow $x_Lp^+$ should then be restricted by 
uncertainty principle. Note that the phase factors in the off-diagonal
matrix elements have an additional factor of $\exp(\pm i x_Lp^+y_2^-)$
relative to their diagonal counterpart. Assuming the same Gaussian nuclear 
distribution in light-cone coordinates and using
\begin{equation}
\int_0^\infty dy^- \rho_0 e^{-{y^-}^2/2{R_A^-}^2 \pm i x_Lp^+y^-}
=e^{-x_L^2/x_A^2}\, ,
\end{equation}
we should have a similar approximation for the off-diagonal matrix elements:
\begin{eqnarray}
\int \frac{dy^{-}}{2\pi}\, dy_1^-dy_2^- 
&& e^{ix_1p^+y^-+ix_2p^+(y_1^--y_2^-) \pm ix_Lp^+y_2^-} \nonumber \\
&\times&\frac{1}{2}\langle A | \bar{\psi}_q(0)\,
\gamma^+\, F_{\sigma}^{\ +}(y_{2}^{-})\, F^{+\sigma}(y_1^{-})\,\psi_q(y^{-})
| A\rangle \theta(-y_2^-)\theta(y^--y_1^-) \nonumber \\
&=&\frac{C}{x_A} f_q^A(x_1)\, x_2f_g^N(x_2)e^{-x_L^2/x_A^2}\, .
\label{eq:off-mx}
\end{eqnarray}
Relative to the diagonal matrix elements, the off-diagonal ones are
suppressed by a factor $\exp(-x_L^2/x_A^2)$. Combining all the four terms
together, we have
\begin{equation}
T^A_{qg}(x,x_L)=\frac{C}{x_A}[f_q^A(x+x_L)\, x_Tf_g^N(x_T)
+f_q^A(x)(x_L+x_T)f_g^N(x_L+x_T)](1-e^{-x_L^2/x_A^2}) \, . \label{eq:matrixf}
\end{equation}
Notice that $\tau_f=1/x_Lp^+$ is the gluon's formation time. Thus,
$x_L/x_A=L_A^-/\tau_f$ with $L_A^-=R_AM/p^+$ being the nuclear size in
the our chosen frame. It is then clear from the above that the interferences
between double-hard and hard-soft scattering cancel exactly the double
scattering contributions for collinear gluon radiation whose formation time
is much larger than the nuclear size. The LPM effect now explicitly
manifests itself via the effective matrix elements in the
gluon radiation spectrum that is induced by double parton scattering.

The diagonal parton matrix elements that arise from hard-soft and double
hard processes are exactly the same as what appear in processes
with large final transverse momentum such as di-jet production studied in 
Ref.~\cite{LQS}. These diagonal matrix elements have an enhancement
linear in nuclear size $R_A$ relative to the single parton distribution
in nuclei as in Eq.~(\ref{eq:t4matrix}). The off-diagonal parton matrix
elements, on the other hand, come from the interference between hard-soft
and double hard processes. Because such parton matrix elements involve
transferring momentum fraction $x_L$ between two nucleons inside a
nucleus, they should be suppressed when $x_L$ is larger than what the
uncertainty principle allows. For a Gaussian nuclear distribution,
these off-diagonal matrix elements are exponentially suppressed for
large $x_L/x_A$ as in Eq.~(\ref{eq:off-mx}). This is why one can 
neglect the interferences between soft and hard rescattering in 
the LQS \cite{LQS} study of processes with large final 
transverse momentum, $\ell_T>> Q^2/A^{1/3}$. However, in our
current study of fragmentation processes, the off-diagonal matrix elements
in the interferences are very important. The interference terms
cancel exactly the normal contributions that contain the diagonal matrix
elements as shown in Eq.~(\ref{eq:matrixf}) for $\ell_T=0$. For small but
finite $\ell_T$, the cancellation is not complete. The effective
parton matrix element (diagonal minus the off-diagonal ones) is then
proportional to $x_L^2/x_A^3$ which depends cubically on the nuclear
size $R_A$. It, however, also vanishes quadratically with $x_L\sim \ell_T^2$.
Therefore, the interferences effectively cut off the transverse momentum
of the radiated gluon at $\ell_T^2\sim Q^2/A^{1/3}$. In terms of the LPM
interference, this means that the formation time of induced gluon 
radiation cannot be larger than the nuclear size $R_A$. Such a
consequence of the LPM interference effect is critical for us to apply
the LQS generalized factorization to the problem of nuclear
modification of fragmentation function in this study.

\subsection{Parton Energy Loss}

So far we have derived the modification to the quark fragmentation functions
and their DGLAP evolution equations. Given the
twist-four parton matrix elements, one can then solve the modified DGLAP 
evolution equations for the modified fragmentation functions. Doing so, 
one can effectively resum all the leading-log corrections at twist-four.
We leave the numerical solutions to future studies. In this paper, we
instead will formulate the effective energy loss of leading quarks.
 
In principle the modification of the fragmentation functions would be the only
experimental effect of induced gluon radiation via multiple scattering. 
One can never directly measure the energy loss of the leading
quark. The net effect of the energy loss is the suppression of leading 
particles on one hand and the enhancement of soft particles on the other, 
leading to the modification of the fragmentation functions. One can then
experimentally characterize the parton energy loss via the momentum transfer 
from large to small momentum regions of the fragmentation functions. We
will discuss this in detail when we numerically calculate the modified
fragmentation functions.

Upon a close examination of Eq.~(\ref{eq:dmod}), we see that 
the first term is the renormalized fragmentation
function in vacuum. The rest are particle production induced by the 
rescattering of the quark through the nuclear medium. In particular,
the last term is particle production from the fragmentation of the gluon
which is induced by the secondary scattering. Such particle production
is at the expense of the energy loss of the leading quark. We can thus
quantify the quark energy loss by the momentum fraction carried by 
the induced gluon,
\begin{eqnarray}
\langle\Delta z_g\rangle(x_B,\mu^2) 
&=& \int_0^{\mu^2}\frac{d\ell_T^2}{\ell_T^2} 
\int_0^1 dz \frac{\alpha_s}{2\pi}
 z\,\Delta\gamma_{q\rightarrow gq}(z,x_B,x_L,\ell_T^2) \nonumber \\
&=&\int_0^{\mu^2}\frac{d\ell_T^2}{\ell_T^2} \int_0^1 dz [1+(1-z)^2]
\frac{T^A_{qg}(x_B,x_L)}{\widetilde{f}_q^A(x_B,\mu_I^2)}
\frac{C_A\alpha_s^2}{N_c(\ell_T^2+\langle k_T^2\rangle)} \, . \label{eq:loss1}
\end{eqnarray}
Using the approximation of the parton correlation functions in 
Eq.~(\ref{eq:matrixf}) and changing the integration variable, we have
\begin{eqnarray}
\langle\Delta z_g\rangle(x_B,\mu^2)& = &
\frac{C_A\alpha_s^2}{N_c\widetilde{f}_q^A(x_B,\mu_I^2)}
\frac{x_B}{x_AQ^2} C \int_0^1 dz \frac{1+(1-z)^2}{z(1-z)}
\int_0^{x_\mu} \frac{dx_L}{x_L(x_L+x_T)}(1-e^{-x_L^2/x_A^2}) \nonumber \\
&\times &[f_q^A(x_B+x_L)\, x_Tf_g^N(x_T)+f_q^A(x_B)(x_L+x_T)f_g^N(x_L+x_T)]\, ,
\label{eq:loss2}
\end{eqnarray}
where $x_\mu=\mu^2/2p^+q^-z(1-z)=x_B/z(1-z)$ if we choose the 
factorization scale as $\mu^2=Q^2$. 

One can numerically evaluate the above in the future with assumed 
parton distributions in nuclei. It is nevertheless interesting to 
discuss the nuclear dependence of the parton energy.
The exponential factor in the above
equation comes from the LPM interferences. It regularizes the 
integration over $x_L$ which is otherwise divergent. 
Combining with the QCD radiation spectrum, it also
limits the integration over $x_L$ to $x_L<x_A$ and gives
$\int dx_L/x_L^2 \sim 1/x_A$. The fractional energy loss
by the quark is, therefore, proportional to
\begin{equation}
\langle\Delta z_g\rangle \sim \frac{C_A \alpha_s^2}{N_c}
\frac{x_B}{x_A^2 Q^2} \; .
\end{equation}
Since $x_A=1/MR_A$, the energy loss thus depends quadratically on the
nuclear size. The extra size dependence comes from the combination of
the QCD radiation spectrum and the modification of the available
phase space in $\ell_T$ or $x_L$ due to the LPM interferences. One also
notes that, though the fractional energy loss is suppressed by $1/Q^2$ 
for a fixed value of $x_B$, the total energy loss, 
$\Delta E=q^-\langle\Delta z_g\rangle$, is not.

\section{Summary and Discussions}

Working in the framework of generalized factorization of higher-twist
parton distributions, we have studied multiple parton scattering in deeply
inelastic $eA$ collisions, in particular the induced gluon radiation and
the resultant modification to the final hadron spectra. We have defined 
modified quark fragmentation functions to take into account the effect of
multiple parton scattering. We have presented a detailed 
derivation of the modified quark fragmentation functions
and their QCD evolution equations to the next-leading-twist. The modification
is shown to depend on twist-four parton matrix elements, both
diagonal and off-diagonal, in nuclei.

Depending on whether the gluon radiation is induced by the secondary
scattering, one can categorize the multiple parton scattering as soft
or hard, according to the fractional momentum carried by the 
secondary parton involved. We have considered both soft and hard 
scattering and their interferences. We have shown that these are
exactly the so-called LPM interference effect. We have demonstrated
that LPM interferences modify the available phase space in the emitted
gluon's momentum. Coupled with the gluon spectrum in QCD, this leads to
the quadratical dependence of the modification of fragmentation functions
or the effective parton energy loss on the nuclear size $R_A$.

The LPM interference effect in this study is critical for applying
the LQS generalized factorization to the fragmentation processes.
Because of the LPM interference effect, the gluon radiation with small
transverse momentum is suppressed. Equivalently, the formation time
of the gluon radiation has to be smaller than the nuclear size $R_A$. 
This requires that the radiated gluon have a minimum transverse
momentum $\ell_T^2 \sim Q^2/A^{1/3}$. For such a value of transverse
momentum, we can still take the leading logarithmic approximation 
$\ell_T^2\ll Q^2$ in the study of fragmentation function in a large
nucleus ($A\gg 1$). Therefore, the applicability of the our study here
is limited to experimental situations in which $Q^2$ must be large enough
such that $Q^2\gg Q^2/A^{1/3}\gg \Lambda_{\rm QCD}^2$.
In the meantime, the resultant nuclear corrections
proportional to $\alpha_s A^{1/3}/\ell_T^2 \sim \alpha_sA^{2/3}/Q^2$
from double parton scattering are still leading for large values
of $Q^2$. This is why the nuclear corrections to the fragmentation
function and the induced parton energy loss depend quadratically on
the nuclear size $R_A$. In the processes with large final transverse 
momentum $\ell_T^2\sim Q^2$ as studied by LQS \cite{LQS}, the leading 
correction is instead proportional to $\alpha_sA^{1/3}/Q^2$.
This is because one can neglect the interference contribution
if $\ell_T^2\gg Q^2/A^{1/3}$.

We have also considered double-quark scattering processes. Though their
contributions to the QCD evolution equations can be neglected as compared
to quark-gluon scattering, they do have a leading-order contribution
which mixes quark and gluon fragmentation functions. Since they involve
quark-antiquark correlations in nuclei, the modification to quark 
and antiquark fragmentation functions will be different. This might
give different modification to the spectra for negative and positive
hadrons as observed in experiments \cite{hermes}

There is currently little information on the twist-four parton
matrix elements in nuclei, especially the off-diagonal ones. We
have only outlined a very crude estimate, assuming factorization
of the two parton correlations in nuclei. This enables us to estimate
the nuclear dependence of the final results. Future work on this is
necessary in order to have any quantitative study of the problem. With
that information, one should be able to numerically solve the QCD evolution
equation for the modified fragmentation functions.

The method we developed here to study the modification of parton
fragmentation functions cannot yet be applied directly to other
processes, such as high-energy $pA$ and $AA$ collisions. One can, 
however, find important implications from the study in this paper.
The $A^{4/3}$ nuclear size dependence of the twist-four parton matrix 
elements relies on the color confinement within the nucleon radius $r_N$.
So the strength of parton correlation as represented by the constant
$C$ in Eq.~(\ref{eq:t4matrix}) is proportional to $r_N$. If we replace
the nuclear target in DIS by a droplet of quark-gluon plasma, then
the integration over $y_1^- - y_2^-$ in Eq.~(\ref{eq:t4matrix}) is
no longer restricted to $r_N$ but rather to the effective parton 
correlation length $1/\mu_D$ with $\mu_D$ being the Debye screening
mass in the quark-gluon plasma. So the modification of the parton
fragmentation function and the effective parton energy loss will be
sensitive to $\mu_D$. If there is a dramatic change in the parton
correlation during the QCD phase transition, one should then expect
a similar behavior in parton energy loss.

\section*{Acknowledgements}

We thank J. Qiu and G. Sterman for helpful discussions. 
This work was supported 
by the Director, Office of Energy Research, Office of High Energy 
and Nuclear Physics, Divisions of Nuclear Physics, of the U.S. 
Department of Energy under Contract No. DE-AC03-76SF00098 
and DE-FG-02-96ER40989 and in part by NSFC under project 19928511.

\section*{Appendix}

In this Appendix we calculate the hard part of gluon radiation processes
associated with multiple quark-gluon scattering in DIS off a nucleus. 
We use the technique of helicity amplitude for high-energy parton
scattering, where one can neglect the transverse recoil induced by the
scattering. The final results will agree with the complete calculation in
the limit of soft radiation. Such an exercise can help us to cross-check 
the results of our complete calculation. More importantly, it can help
us to understand the physical processes more clearly on the level of
amplitudes and to reorganize the final results in this paper 
according to the nature of different processes.

\subsection{Single Scattering}

Assuming the dominant component of a fast
quark is $\ell_q\approx[0,\ell_q^-,\vec{0}_\perp]$. One has,
\begin{equation}
\bar{u}(\ell_q,\lambda)\gamma^\mu u(\ell_q',\lambda')
\approx 2\sqrt{\ell_q^-\ell_q'^-} \,\, 
\delta_{\lambda\lambda'}\,\underline{n}^\mu\, , \label{eq:heli}
\end{equation}
where $\underline{n}=[0,1,\vec{0}_\perp]$, $\lambda$ and $\lambda'$ are
quark helicities. According to the collinear factorization of the
parton matrix elements, each initial quark line contributes
$u(p)$ while each initial gluon line (with Lorentz 
index $\sigma$) contributes $p_\sigma$ to the amplitude. Each initial
parton line with momentum $p_i=[x_ip^+,0,\vec{k}_T]$ also has a momentum
integral $dx_i/2\pi$ with a phase factor 
$\exp(-ix_ip^+y_i^- +i\vec{k}_T\cdot\vec{y}_{Ti})$. We summarize
these special Feynman rules for initial parton lines as:
\begin{eqnarray}
{\rm initial\,\, quark} &\rightarrow& u(p)
\int \frac{dx_i}{2\pi} e^{-ix_ip^+y_i^- +i\vec{k}_T\cdot\vec{y}_{Ti}},
\nonumber \\ 
{\rm initial\,\, gluon} &\rightarrow& p_\sigma 
\int \frac{dx_i}{2\pi} e^{-ix_ip^+y_i^- +i\vec{k}_T\cdot\vec{y}_{Ti}}.
\end{eqnarray}
All the internal and final external parton lines follow 
the normal Feynman rules.

We will work with an axial gauge $A^-=0$. The final gluon with momentum 
$\ell$ is assumed to carry $1-z$ momentum fraction of the struck quark 
and has polarization $\varepsilon(\ell)$,
\begin{eqnarray}
\ell&=&\left[\frac{\ell_T^2}{2(1-z)q^-}, 
(1-z)q^-, \vec{\ell_T}\right], \nonumber \\
\varepsilon(\ell)&=&\left[\frac{\vec{\epsilon}_T\cdot\vec{\ell}_T}{(1-z)q^-},
0,\vec{\epsilon}_T\right]\, .
\end{eqnarray}

\begin{figure}
\centerline{\psfig{file=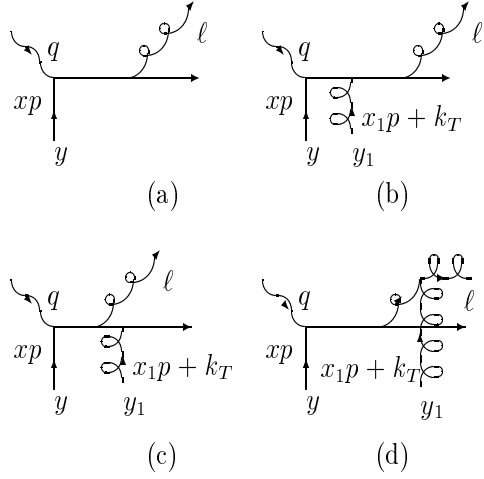,width=2.5in,height=2.5in}}
\caption{Gluon radiation from a single scattering (a) and double
scattering (b-d) in DIS.}
\label{fig:sd}
\end{figure}

According to the above Feynman rules, the amplitude for gluon radiation
in the single scattering case of DIS as shown in Fig.~\ref{fig:sd}(a) is
\begin{eqnarray}
M_\mu^S(y)&=&\int\frac{dx}{2\pi}
\bar{u}(xp+q)\gamma_\mu u(p) \overline{M}^S \, , \nonumber \\
\overline{M}^S(y)&=&
2g\frac{\vec{\epsilon}_T\cdot\vec{\ell}_T}{\ell_T^2} 
T_c e^{-i(x_B+x_L)p^+y^-}\, ,
\end{eqnarray}
where we have used the on-shell condition for the final quark
\begin{equation}
  (xp+q-\ell)^2=2p^+q^-(x-x_B-x_L)=0 \, ,
\end{equation}
and assumed soft radiation limit $1-z\rightarrow 0$. $T_c$ is the color
matrix in the adjunct representation with the color index $c$ for the
radiated gluon.

To calculate the hard part of the parton scattering, one can simply take the
square of the amplitude $M_\nu^{S(1)}(0){M_\mu^{S(1)}}^\dagger(y)$, 
average over the initial quark's spin, sum over
the final gluon's polarization and integrate over the final partons'
momenta. One should also average over initial and sum over the final
partons' color indices. 
Note that one of integrations over $x$ is carried out by the
overall momentum conservation $2\pi\delta(\sum x_i)$. In the other integral
$dx/2\pi$, we absorb the factor $1/2\pi$ into the definition of the parton
matrix elements associated with the processes. Following this convention,
we have the hard part of gluon radiation induced by single scattering,
\begin{eqnarray}
\overline{H}^{(1)}_{\mu\nu}&=& \int dx 
\frac{1}{2}{\rm Tr}[p\cdot\gamma\gamma_\mu (xp+q)\cdot\gamma \gamma_\nu]
2\pi\frac{\delta(x-x_B)}{2p^+q^-} \nonumber \\
&\times&\int\frac{d^2\ell_T}{2(2\pi)^3}\frac{dz}{1-z}
\overline{M}^S(0){\overline{M}^S}^\dagger(y)
\nonumber \\
&=&\int dx H^{(0)}_{\mu\nu}(x,p,q) \frac{\alpha_s}{2\pi}
\int\frac{d\ell_T^2}{\ell_T^2} dz C_F \frac{2}{1-z} e^{i(x+x_L)p^+y^-}\, .
\end{eqnarray}
One can see that the above is the same as the hard part in
Eq.~(\ref{eq:h11}) in the soft radiation limit $z\rightarrow 1$.   

\subsection{Double Scattering}

For double scattering processes as shown in Fig.~\ref{fig:sd}(b)-(d),
there is one loop-integration which is carried out by the contour
integration around a pole in one of the two propagators. We also
assume that the initial gluon carries transverse momentum $k_T$. The on-shell
requirement of the final quark is
\begin{equation}
[q+(x+x_1)p+k_T-\ell]^2=2zp^+q^-(x+x_1-x_B-x_L-x_D)=0 \; ,
\end{equation}
where $x_L$ and $x_D$ are defined in Eq.~(\ref{eq:xld}). The kinematics
in the process of Fig.~\ref{fig:sd}(b) only allows one choice of
pole. The contour integration around this pole gives
\begin{equation}
\int \frac{dx}{2\pi}\frac{e^{-ixp^+(y^--y_1^-)-i(x_B+x_L+x_D)p^+y_1^-}}
{(xp+q)^2+i\epsilon} =\frac{i}{2p^+q^-}
e^{-ix_Bp^+y^- -i(x_L+x_D)p^+y_1^-}\theta(y^--y_1^-)\, .
\end{equation}
The momentum fraction carried by the initial gluon is 
then $x_1=x_L+x_D$. Since this momentum fraction is finite when $k_T=0$, 
the secondary scattering is considered hard. The final gluon in this process
is from the final state radiation of the secondary scattering. 
This is what we refer to as double hard scattering.
Using the helicity selection rule in Eq.~(\ref{eq:heli}), one can find the
amplitude for the process in Fig.~\ref{fig:sd}(b),
\begin{eqnarray}
\overline{M}^{D(1)}(y,y_1)&=&
2g\frac{\vec{\epsilon}_T\cdot\vec{\ell}_T}{\ell_T^2} 
T_cT_{a_1} e^{-i(x_B+x_L)p^+y^- -ix_Dp^+y_1^-}\nonumber \\
&\times&
e^{ix_Lp^+(y^--y_1^-)} ig \theta(y^--y_1^-)\, ,
\end{eqnarray}
where $a_1$ is the color index of the initial gluon. Note that
there should also be an integral over $k_T$. We have absorbed this
together with the transverse phase factor 
$\exp(i\vec{k_T}\cdot\vec{y}_{1T})$ into the definition of parton matrix
elements.

In the other two diagrams of double scattering
in Fig.~\ref{fig:sd}, there are two possible
choices of poles. One choice of poles corresponds to soft rescattering
of the final quark [Fig.~\ref{fig:sd}(c)] or of the final 
gluon [Fig.~\ref{fig:sd}(d)] where the initial gluon's momentum fraction
$x_1=x_D$ [for Fig.~\ref{fig:sd}(c)] or $x_1=-zx_D/(1-z)$ 
[for Fig.~\ref{fig:sd}(d)] goes to zero for $k_T=0$. The final gluon
is induced by the hard photon-quark scattering. This is
what we call hard-soft double scattering. The second choice however 
gives the initial gluon finite momentum $x_1=x_L+x_D$ and the 
gluon radiation is from the initial state radiation of the hard 
secondary scattering in what we call double-hard scattering. 
One can find the amplitudes of these two diagrams as
\begin{eqnarray}
\overline{M}^{D(2)}(y,y_1)&=&
2g\frac{\vec{\epsilon}_T\cdot\vec{\ell}_T}{\ell_T^2} 
T_{a_1}T_c e^{-i(x_B+x_L)p^+y^- -ix_Dp^+y_1^-}\nonumber \\
&\times&
[1-e^{ix_Lp^+(y^--y_1^-)}] ig \theta(y^--y_1^-)\, , \\
\overline{M}^{D(3)}(y,y_1)&=&
-2g\frac{\vec{\epsilon}_T\cdot(\vec{\ell}_T-\vec{k}_T)}
{(\vec{\ell}_T-\vec{k}_T)^2}
[T_{a_1},T_c] e^{-i(x_B+x_L)p^+y^- -ix_Dp^+y_1^-}\nonumber \\
&\times&
[e^{-ix_Dp^+(y^--y_1^-)/(1-z)}-e^{ix_Lp^+(y^--y_1^-)}] ig \theta(y^--y_1^-)\, .
\end{eqnarray}
Since the gluon radiation induced by the hard rescattering in these two 
diagrams is initial state radiation, the amplitude has the opposite sign
with respect to the hard-soft processes where the gluon is produced by
final state radiation of the first hard scattering.

\subsection{Triple Scattering}
\begin{figure}
\centerline{\psfig{file=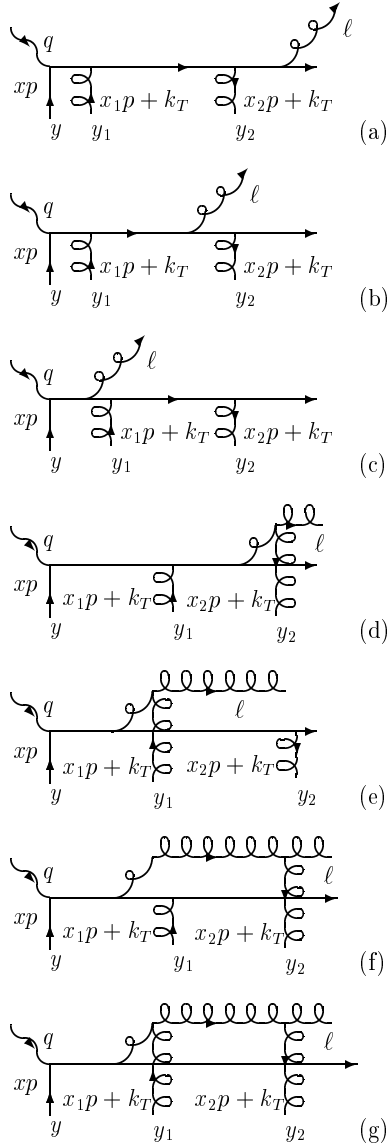,width=2.0in,height=6.0in}}
\caption{Gluon radiation from triple scattering in DIS.}
\label{fig:triple}
\end{figure}

For gluon radiation associated with triple scattering, there are all
together 7 different diagrams as shown in Fig.~\ref{fig:triple}. we
list the amplitudes here:

\begin{eqnarray}
\overline{M}^{T(1)}(y,y_1,y_2)&=&
2g\frac{\vec{\epsilon}_T\cdot\vec{\ell}_T}{\ell_T^2} 
T_cT_{a_2}T_{a_1} e^{-i(x_B+x_L)p^+y^- -ix_D^0p^+(y_1^--y_2^-)}\nonumber \\
&\times&
e^{ix_Lp^+(y^--y_2^-)} (-g^2) \theta(y_1^--y_2^-)\theta(y^--y_1^-)\, , 
\end{eqnarray}
\begin{eqnarray}
\overline{M}^{T(2)}(y,y_1,y_2)&=&
2g\frac{\vec{\epsilon}_T\cdot\vec{\ell}_T}{\ell_T^2}
T_{a_2}T_cT_{a_1} e^{-i(x_B+x_L)p^+y^- -ix_Dp^+(y_1^--y_2^-)}\nonumber \\
&\times&
[e^{ix_Lp^+(y^--y_1^-)}
-e^{ix_Lp^+(y^--y_2^-)-i(x_D^0-x_D)p^+(y_1^--y_2^-)}] \nonumber \\
&\times&(-g^2) \theta(y_1^--y_2^-)\theta(y^--y_1^-)\, ,
\end{eqnarray}
\begin{eqnarray}
\overline{M}^{T(3)}(y,y_1,y_2)&=&
2g\frac{\vec{\epsilon}_T\cdot\vec{\ell}_T}{\ell_T^2} 
T_{a_2}T_{a_1}T_c e^{-i(x_B+x_L)p^+y^- -ix_Dp^+(y_1^--y_2^-)}\nonumber \\
&\times&
[1-e^{ix_Lp^+(y^--y_1^-)}] (-g^2) \theta(y_1^--y_2^-)\theta(y^--y_1^-)\, , 
\end{eqnarray}

\begin{eqnarray}
\overline{M}^{T(4)}(y,y_1,y_2)&=&
2g\frac{\vec{\epsilon}_T\cdot(\vec{\ell}_T-\vec{k}_T)}
{(\vec{\ell}_T-\vec{k}_T)^2} 
[T_{a_2},T_c]T_{a_1}e^{-i(x_B+x_L)p^+y^- -ix_Dp^+(y_1^--y_2^-)}\nonumber \\
&\times&
[e^{-i(x_D^0-x_D)p^+(y_1^--y_2^-)+ix_Lp^+(y^--y_2^-)}\nonumber \\
&-&e^{i(1-z/(1-z))x_Dp^+(y_1^--y_2^-)+ix_Lp^+(y^--y_1^-)}] 
(-g^2) \theta(y_1^--y_2^-)\theta(y^--y_1^-)\, , 
\end{eqnarray}
\begin{eqnarray}
\overline{M}^{T(5)}(y,y_1,y_2)&=&
2g\frac{\vec{\epsilon}_T\cdot(\vec{\ell}_T-\vec{k}_T)}
{(\vec{\ell}_T-\vec{k}_T)^2} 
T_{a_2}[T_{a_1},T_c]e^{-i(x_B+x_L)p^+y^- -ix_Dp^+(y_1^--y_2^-)}\nonumber \\
&\times&
[e^{ix_Lp^+(y^--y_1^-)}
-e^{-ix_Dp^+(y^--y_1^-)/(1-z)}]
(-g^2) \theta(y_1^--y_2^-)\theta(y^--y_1^-)\, , 
\end{eqnarray}
\begin{eqnarray}
\overline{M}^{T(6)}(y,y_1,y_2)&=&
2g\frac{\vec{\epsilon}_T\cdot(\vec{\ell}_T-\vec{k}_T)}
{(\vec{\ell}_T-\vec{k}_T)^2} 
T_{a_1}[T_{a_2},T_c]e^{-i(x_B+x_L)p^+y^- -ix_Dp^+(y_1^--y_2^-)}\nonumber \\
&\times& \left[e^{ix_Lp^+(y^--y_1^-)}
-e^{-ix_Dp^+(y^--y_1^-)/(1-z)}\right] e^{i(1-z/(1-z))x_Dp^+(y_1^--y_2^-)}
\nonumber \\
&\times&(-g^2) \theta(y_1^--y_2^-)\theta(y^--y_1^-)\, , 
\end{eqnarray}
\begin{eqnarray}
\overline{M}^{T(7)}(y,y_1,y_2)&=&
2g\frac{\vec{\epsilon}_T\cdot\vec{\ell}_T}
{\vec{\ell}_T^2} 
[T_{a_1},[T_{a_2},T_c]]e^{-i(x_B+x_L)p^+y^- -ix_Dp^+(y_1^--y_2^-)}\nonumber \\
&\times&
[1-e^{ix_Lp^+(y^--y_1^-)}]
e^{i(1-z/(1-z))x_Dp^+(y_1^--y_2^-)} \nonumber \\
&\times&(-g^2) \theta(y_1^--y_2^-)\theta(y^--y_1^-)\, .
\end{eqnarray}
In $\overline{M}^{T(4)}$, $\overline{M}^{T(6)}$ and $\overline{M}^{T(7)}$,
where the radiated gluon is coupled to the second initial gluon through
a triple-gluon vertex, we have made variable change $k_T\rightarrow -k_T$. 

To obtain these amplitudes, two loop-integrations are carried out
by contour integrations around two poles. For each diagram, there
are sometimes two possible choices of the two poles, and consequently
two contributions to the amplitude. They represent soft or hard
rescattering processes. However, one of the quark-gluon scatterings
must be soft. In Fig.~\ref{fig:triple}(a), for example, the
gluon induced by the third (hard) scattering after the quark has a
soft second scattering. Fig~\ref{fig:triple}(b) has two contributions.
One corresponds to gluon radiation induced by the third scattering,
the other by the second scattering. One of the two quark-gluon scatterings
must be soft. In  Fig~\ref{fig:triple}(c), one contribution comes
from gluon radiation induced by the second scattering followed by
soft quark-gluon scattering. The other contribution corresponds
to gluon radiation from the first photon-quark hard scattering and
double soft scattering afterwards. Such analyses of each diagram help
us to reorganize the amplitude according to the physical processes.
 
\subsection{Soft and Hard Rescattering}
We can reorganize all the radiation amplitudes of double and triple scattering
according to our classification of soft and hard rescattering. The total
amplitude for the hard-soft double scattering in Figs.~\ref{fig:sd}(b)-(d),
as represented by Fig.~\ref{fig7}, is
\begin{eqnarray}
\overline{M}^D_S(y,y_1)&=&
e^{-i(x_B+x_L)p^+y^- -ix_Dp^+y_1^-} ig \theta(y^--y_1^-) \nonumber \\
&\times&2g\left[\frac{\vec{\epsilon}_T\cdot\vec{\ell}_T}{\ell_T^2}T_{a_1}T_c
-\frac{\vec{\epsilon}_T\cdot(\vec{\ell}_T-\vec{k}_T)}
{(\vec{\ell}_T-\vec{k}_T)^2}[T_{a_1},T_c]
e^{-ix_Dp^+(y^--y_1^-)/(1-z)}\right] \, .
\end{eqnarray}
The amplitude for double hard scattering, 
as represented by Fig.~\ref{fig8}, is,
\begin{eqnarray}
\overline{M}^D_H(y,y_1)&=&
e^{-i(x_B+x_L)p^+y^- -ix_Dp^+y_1^-} ig \theta(y^--y_1^-) \nonumber \\
&\times&2g[T_c,T_{a_1}]
\left[ \frac{\vec{\epsilon}_T\cdot\vec{\ell}_T}{\ell_T^2}
-\frac{\vec{\epsilon}_T\cdot(\vec{\ell}_T-\vec{k}_T)}
{(\vec{\ell}_T-\vec{k}_T)^2}\right]e^{ix_Lp^+(y^--y_1^-)}\, .
\end{eqnarray}
In triple scattering processes, the gluon radiation can be induced by
the initial photon-quark hard scattering which is then followed by two
soft scatterings. We denote the gluon radiation together with the first soft 
quark-gluon or gluon-gluon scattering by the effective diagram in 
Fig.~\ref{fig7}. Similarly, the second soft scattering can be either 
quark-gluon or gluon-gluon as shown in Fig.~\ref{fig10}(d) and (e). The 
amplitudes for these two diagrams with double soft scattering are,
\begin{eqnarray}
\overline{M}^T_{S(1)}(y,y_1,y_2)&=&
e^{-i(x_B+x_L)p^+y^- -ix_Dp^+(y_1^--y_2^-)}
(-g^2) \theta(y_1^--y_2^-)\theta(y^--y_1^-) \nonumber \\
&\times&2g\left[\frac{\vec{\epsilon}_T\cdot\vec{\ell}_T}{\ell_T^2} 
T_{a_2}T_{a_1}T_c
-\frac{\vec{\epsilon}_T\cdot(\vec{\ell}_T-\vec{k}_T)}
{(\vec{\ell}_T-\vec{k}_T)^2} T_{a_2}[T_{a_1},T_c]
e^{-ix_Dp^+(y^--y_1^-)/(1-z)} \right]\, ,
\end{eqnarray}
\begin{eqnarray}
\overline{M}^T_{S(2)}(y,y_1,y_2)&=&
e^{-i(x_B+x_L)p^+y^- -ix_Dp^+(y_1^--y_2^-)}
(-g^2) \theta(y_1^--y_2^-)\theta(y^--y_1^-) \nonumber \\
&\times&2g\left[-\frac{\vec{\epsilon}_T\cdot(\vec{\ell}_T-\vec{k}_T)}
{(\vec{\ell}_T-\vec{k}_T)^2} T_{a_1}[T_{a_2},T_c]
e^{-ix_Dp^+(y^--y_1^-)/(1-z)}\right. \nonumber \\
&+&\left.\frac{\vec{\epsilon}_T\cdot\vec{\ell}_T}{\ell_T^2} 
[T_{a_1},[T_{a_2},T_c]]
\right]e^{i(1-z/(1-z))x_Dp^+(y_1^--y_2^-)}
\, ,
\end{eqnarray}
where $\overline{M}^T_{S(1)}$ is the combination of the first term in
$\overline{M}^{T(3)}$ and the second term in $\overline{M}^{T(5)}$ while
$\overline{M}^T_{S(2)}$ is the combination of the second term of 
$\overline{M}^{T(6)}$ and the first term of $\overline{M}^{T(7)}$.

In hard rescattering, the gluon radiation is induced  either by the
second or the third scattering. The amplitude for induced radiation
by the third scattering as shown in Fig.~\ref{fig10}(c) is
\begin{eqnarray}
\overline{M}^T_{H(1)}(y,y_1,y_2)&=&
e^{-i(x_B+x_L)p^+y^- -ix_D^0p^+(y_1^--y_2^-)}
(-)g^2\theta(y_1^--y_2^-)\theta(y^--y_1^-)   \nonumber \\
&\times&2g[T_c,T_{a_2}]T_{a_1}
\left[ \frac{\vec{\epsilon}_T\cdot\vec{\ell}_T}{\ell_T^2}
-\frac{\vec{\epsilon}_T\cdot(\vec{\ell}_T-\vec{k}_T)}
{(\vec{\ell}_T-\vec{k}_T)^2}\right]e^{ix_Lp^+(y^--y_2^-)}\, .
\end{eqnarray}
It is the combination of $\overline{M}^{T(1)}$, the second term in
$\overline{M}^{T(2)}$, and the first term in $\overline{M}^{T(4)}$.
If the gluon is induced by the second scattering, the third soft
scattering can either be quark-gluon or gluon-gluon as shown in
Figs.~\ref{fig10}(a) and (b). Their amplitudes are
\begin{eqnarray}
\overline{M}^T_{H(2)}(y,y_1,y_2)&=&
e^{-i(x_B+x_L)p^+y^- -ix_Dp^+(y_1^--y_2^-)}
(-)g^2\theta(y_1^--y_2^-)\theta(y^--y_1^-)   \nonumber \\
&\times&2gT_{a_2}[T_c,T_{a_1}]
\left[ \frac{\vec{\epsilon}_T\cdot\vec{\ell}_T}{\ell_T^2}
-\frac{\vec{\epsilon}_T\cdot(\vec{\ell}_T-\vec{k}_T)}
{(\vec{\ell}_T-\vec{k}_T)^2}\right]e^{ix_Lp^+(y^--y_1^-)}\, ,
\end{eqnarray}
\begin{eqnarray}
\overline{M}^T_{H(3)}(y,y_1,y_2)&=&
e^{-i(x_B+x_L)p^+y^- -ix_Dp^+(y_1^--y_2^-)}
(-)g^2\theta(y_1^--y_2^-)\theta(y^--y_1^-)   \nonumber \\
&\times&2g[T_{a_1},[T_c,T_{a_2}]]
\left[ \frac{\vec{\epsilon}_T\cdot\vec{\ell}_T}{\ell_T^2}
-\frac{\vec{\epsilon}_T\cdot(\vec{\ell}_T-\vec{k}_T)}
{(\vec{\ell}_T-\vec{k}_T)^2}\right]
e^{ix_Lp^+(y^--y_1^-)} \nonumber \\
&\times&e^{i(1-z/(1-z))x_Dp^+(y_1^--y_2^-)}\, ,
\end{eqnarray}
respectively. Again, $\overline{M}^T_{H(2)}$ is the combination of
the first term in $\overline{M}^{T(2)}$, the second term in
$\overline{M}^{T(3)}$ and the first term of $\overline{M}^{T(5)}$.
$\overline{M}^T_{H(3)}$ is the sum of the second term in 
$\overline{M}^{T(4)}$, the first term of $\overline{M}^{T(6)}$
and the second term of $\overline{M}^{T(7)}$.

With the amplitudes of the re-classified diagrams, we can easily
calculate the partonic hard part of double scattering,
\begin{equation}
\overline{H}^D=
\int\frac{d^2\ell_T}{2(2\pi)^3}\frac{dz}{1-z}
[\overline{M}^D(0,y_2){\overline{M}^D}^\dagger(y,y_1)
+\overline{M}^T(0,y_2,y_1){\overline{M}^S}^\dagger(y)
+\overline{M}^S(0){\overline{M}^T}^\dagger(y,y_1,y_2)]\, ,
\end{equation}
which includes both the double scattering and the interferences of
triple and single scattering. The final results are the same as listed
in Sec.~\ref{sec:list} with the splitting functions replaced by their
form in the limit of $z\rightarrow 1$.

%\begin{multicols}{2}

%\end{multicols}

\end{document}